\documentclass[aps,superscriptaddress,nofootinbib,showpacs]{revtex4}
\usepackage{epsfig,graphics,amssymb,amsmath,subeqnarray,color}

\begin{document}

\title{The geometry and wetting of capillary folding}
\author{Jean-Philippe P\'eraud}
\email{jperaud@mit.edu}
\affiliation{Department of Mechanical and Aerospace Engineering, 
University of California San Diego, 
9500 Gilman Drive, La Jolla CA 92093-0411, USA.}
\affiliation{Department of Mechanical Engineering, Massachusetts Institute of Technology, Cambridge MA 02139, USA.}
\author{Eric Lauga}
\email{e.lauga@damtp.cam.ac.uk}
\affiliation{Department of Mechanical and Aerospace Engineering, 
University of California San Diego, 
9500 Gilman Drive, La Jolla CA 92093-0411, USA.}
\affiliation{Department of Applied Mathematics and Theoretical Physics,  University of Cambridge, 
Centre for Mathematical Sciences, Wilberforce Road, Cambridge, CB3 0WA, United Kingdom.}
\date{\today}

\begin{abstract}
Capillary forces are involved in a variety of natural phenomena, ranging from droplet breakup to the physics of clouds. The forces from surface tension can also be exploited in industrial application provided the length scales involved are small enough. Recent experimental investigations showed how to take advantage of capillarity to fold planar structures into three-dimensional configurations by selectively melting polymeric hinges joining otherwise rigid shapes. In this paper we use theoretical calculations to quantify the role of geometry and fluid wetting on the final folded state. Considering folding in two and three dimensions, studying both hydrophilic and hydrophobic situations with possible contact angle hysteresis, and  addressing the shapes to be folded to be successively infinite, finite, curved, kinked, elastic, we are able to derive an overview of the geometrical  parameter space available for capillary folding. 
\end{abstract}

\maketitle

\section{Introduction}

Capillarity, a field at the intersection of fluid mechanics and statistical physics \cite{degennes85} and already over 200 year old \cite{pomeau06}, not only strikes by its  beauty but also by its broad range of application. It is  involved in  fundamental physical problems including the shape of minimal surfaces, the physics of clouds, and the dynamics of complex fluids \cite{degennes_book}  and  plays also an important role in a variety of  applied processes such as spray formation, foam dynamics, and industrial coatings \cite{pomeau06}. At the scale of the capillary length, typically on the order of millimeters,  surface tension dominates gravitational forces,  resulting in a number of commonly observed phenomena such as the bundling of wet hair at the swimming pool  \cite{bico04},  possibilities for the control of fluid transport  \cite{huang13} and the design of  devices for filtering or sensing \cite{squires05}.

One aspect in which capillary forces could be particularly useful  is a means to apply small forces and deform flexible objects at the microscale, possibly leading to  new manufacturing techniques based on capillary-driven self-assembly of objects. In the submillimeter range of length scales, manufacturing three-dimensional objects remains a challenge. Classical processes such as photolithography are, by essence, two-dimensional \cite{leong07}. One concept who has long been discussed and implemented consists in cleverly patterning two-dimensional structures and then folding them into three-dimensional objects. 

The idea of exploiting surface tension for folding purposes has been proposed and explored in a number of  studies. In the case of smooth surfaces \cite{roman2010elasto}, Py and co-authors \cite{py07,py09}  studied the deformations induced by droplets of liquid deposited on flat flexible sheets with specified shapes subsequently leading to folding (so-called capillary origami). They also  determined numerically  the folding geometry for two-dimensional  elastic sheets in the particular case where the contact line of the droplet remains pinned on the edge of the sheet. A dynamic version of the same idea where the fluid droplet   impacts the flexible structure was  suggested   \cite{antkowiak11}. Surface tension can also be used to induce buckling of an elastic rod  \cite{neukirch06}  and further investigations  implemented the droplet evaporation concept on templates with well-defined hinges   \cite{van2010elastocapillary}.

\begin{figure}[t]
\includegraphics[width=.8\textwidth]{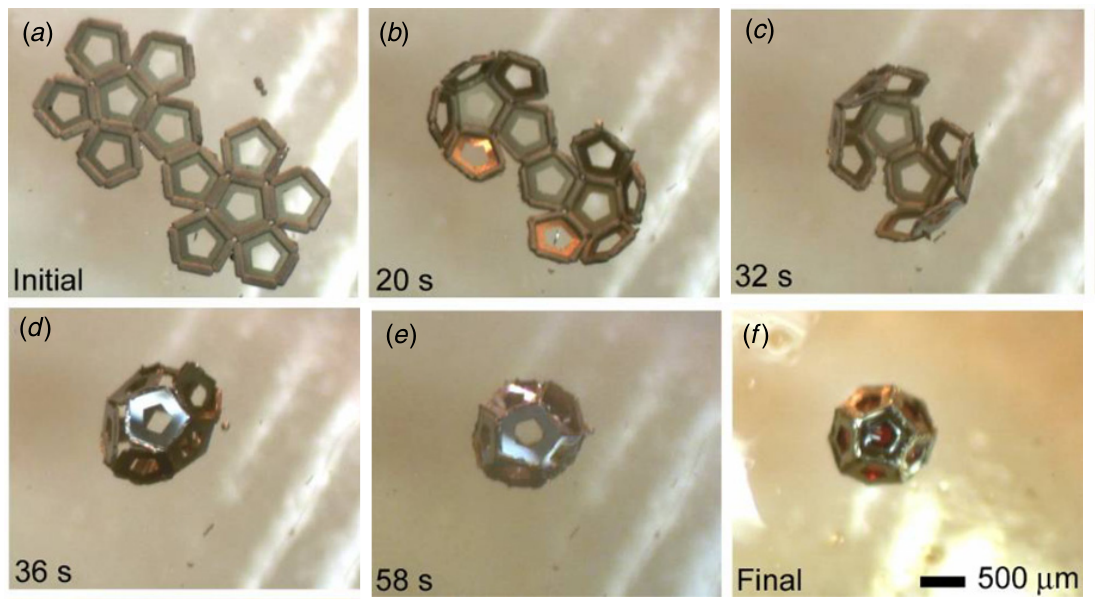}
\caption{Experimental  self-assembly 
of a 500-$\mu$m metallic  dodecahedron using surface tension-induced folding from   molten hinges. Image reproduced from Ref.~\cite{filipiak2009hierarchical} with permission. Copyright 2009 IOP Publishing.}
\label{exp}
\end{figure}

Beyond the work on elastic sheets, capillary forces were also exploited to enable precise micro-manufacturing and three-dimensional encapsulation, an approach pioneered by the Gracias group at Johns Hopkins University to assemble polyhedra \cite{leong07}. The idea is to first design two-dimensional templates using traditional photolithography with pre-selected hinges and faces of different materials. In a second step, polymeric materials located on the hinges (solder) are selectively melted and the resulting capillary forces transform  the two-dimensional template into a three-dimensional folded structure.  This process is illustrated in Fig.~\ref{exp} in the case of metallic dodecahedron (reproduced from Ref.~\cite{filipiak2009hierarchical} with permission).
The early work designed metallic cubes from a  cruciform-shaped template \cite{leong07}. Subsequent work induced folding using  thin-film stress-driven mechanism with no capillary forces \cite{leong08}. A large variety of shapes were also obtained using patterning on several layers of materials \cite{bassik08}. By including a soluble sacrificial layer and  carefully designing the mechanical properties of the other layers a range of three-dimensional structures can be engineered \cite{bassik08}. More recent work was able to generate complex polyhedra \cite{filipiak2009hierarchical}, decrease the length scales involved below microns  \cite{cho2009self}, and create purely polymeric folded structures  \cite{azam2011self}. Ultimately, surface tension may provide a robust mechanism to fold complex three-dimensional structures with important potential biomedical applications, including drug delivery, cell encapsulation, and the fabrication of biocompatible materials with well-defined porosity (see Refs.~\cite{ionov2011soft,randall2012self,fernandes2012self,crane2013fluidic} and references therein). 

From a theoretical standpoint, few models have been developed to quantify  the geometry and dynamics of folding by surface tension. The original capillary origami paper considered the balance between surface and bending energies in the case of complete wetting by the fluid on the solid surface, putting forward an important elasto-capillary length \cite{py07}.  Further work considered the two-dimensional bending of flexible strips and the resulting encapsulation phenomena that may occur, accounting for gravity of the sheet or evaporation of the fluid, but in both cases assuming the droplet contact line to be pinned \cite{rivetti12}. In the case of polyhedra folding using fluidic hinges, numerical simulations using Surface Evolver \cite{brakke92} showed that  the  folding angle depends strongly on the solder volume \cite{leong07}.

In the current paper, we propose to explore the parameter space for folding by focusing on the equilibrium geometry of simple configurations as building blocks for more complex three-dimensional structures.
The emphasis is on the main control parameters for folding, namely (a) the volume of the drop, which could be controlled either by active change or by passive change due to drying or condensation; (b) the surface energies, directly linked to the equilibrium contact angle through Young's equation, with possible contact angle hysteresis;  (c) the geometry of the structures to be folded; and (d)  the flexibility of the solid surface when relevant. We assume to be below the capillary length and ignore the effect of gravity. In addition, most of our study focuses on two-dimensional geometries with some  extensions to three dimensions.

This article is organized as follows. In \S \ref{2D}, we present a series of idealized two-dimensional models in which we calculate the equilibrium configuration of two rigid walls folded by a droplet and vary a number of the model parameters (contact angle, size and geometry of the walls, number of droplets,  and possible de-wetting near the hinge). In \S \ref{hyst} we address the case in which the contact line between the droplet and the solid surface displays hysteresis,  showing in particular how hysteresis can lead to folded structures whose shapes depend on  initial configurations and 
may also help to prevent  de-wetting near hinge points. In \S \ref{3D} we generalize  the two-dimensional   results from  \S \ref{2D} to  three dimensions. This is followed in \S \ref{elastic} by a theoretical study of the two-dimensional folding of an elastic sheet by a droplet with a free contact line by focusing on the role of wetting and droplet size. Our results are finally summarized in \S\ref{discussion}.

\section{Folding in two dimensions}
\label{2D}

In this section, we study the system composed of two freely-hinged rigid walls subject to  capillary forces only from a droplet in the absence of contact angle hysteresis. Starting from the  simplest configuration with one droplet and   two planar infinite walls, we progressively add to more complex features (finite walls, curved walls, two droplets, de-wetting effects) in order to assess the influence of the different geometrical and physical parameters. We demonstrate in particular the importance of the equilibrium contact angle and  the droplet  volume  on the final folding angle.

\subsection{Idealized folding using a single droplet}
\label{single_droplet}
\begin{figure}[t]
\includegraphics[width=.4\textwidth]{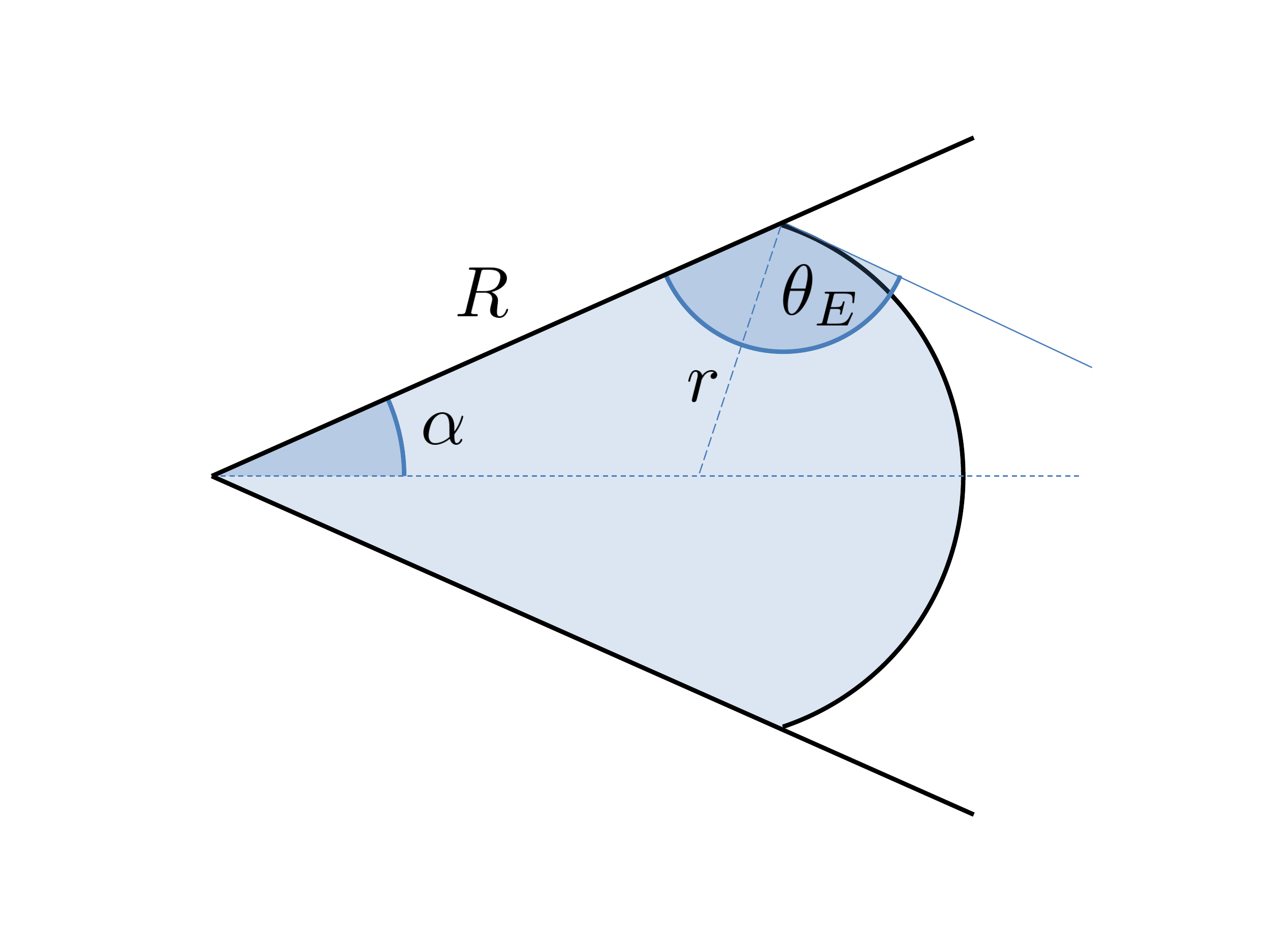}
\caption{
Idealized folding in two dimensions. A fluid droplet is located at the intersection of two freely-hinged walls, opened by a semi-angle $\alpha$ (``the folding angle''). The equilibrium contact  angle  of the droplet on the solid is denoted $\theta_{E}$, and we assume that there is no contact-angle hysteresis (that assumption is relaxed in  \S\ref{hyst}). The length of the wetted region is denoted $R$ and the radius of curvature of the droplet $r$.
}
\label{fig:coin_2D}
\end{figure}

\begin{figure}[t]
\begin{minipage}{.48\textwidth}
\includegraphics[width=\textwidth]{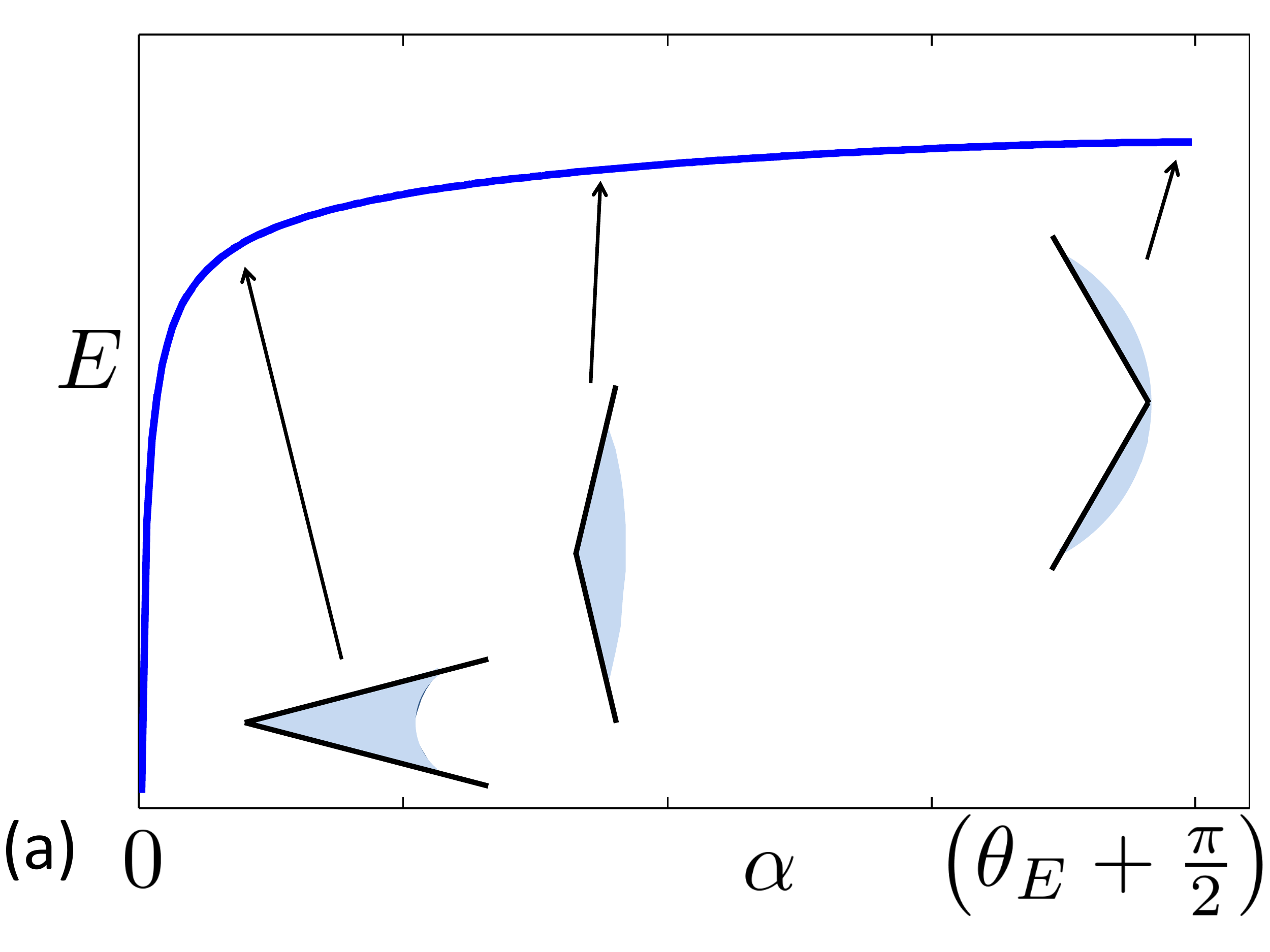}
\end{minipage}\quad\quad
\begin{minipage}{.47\textwidth}
\includegraphics[width=\textwidth]{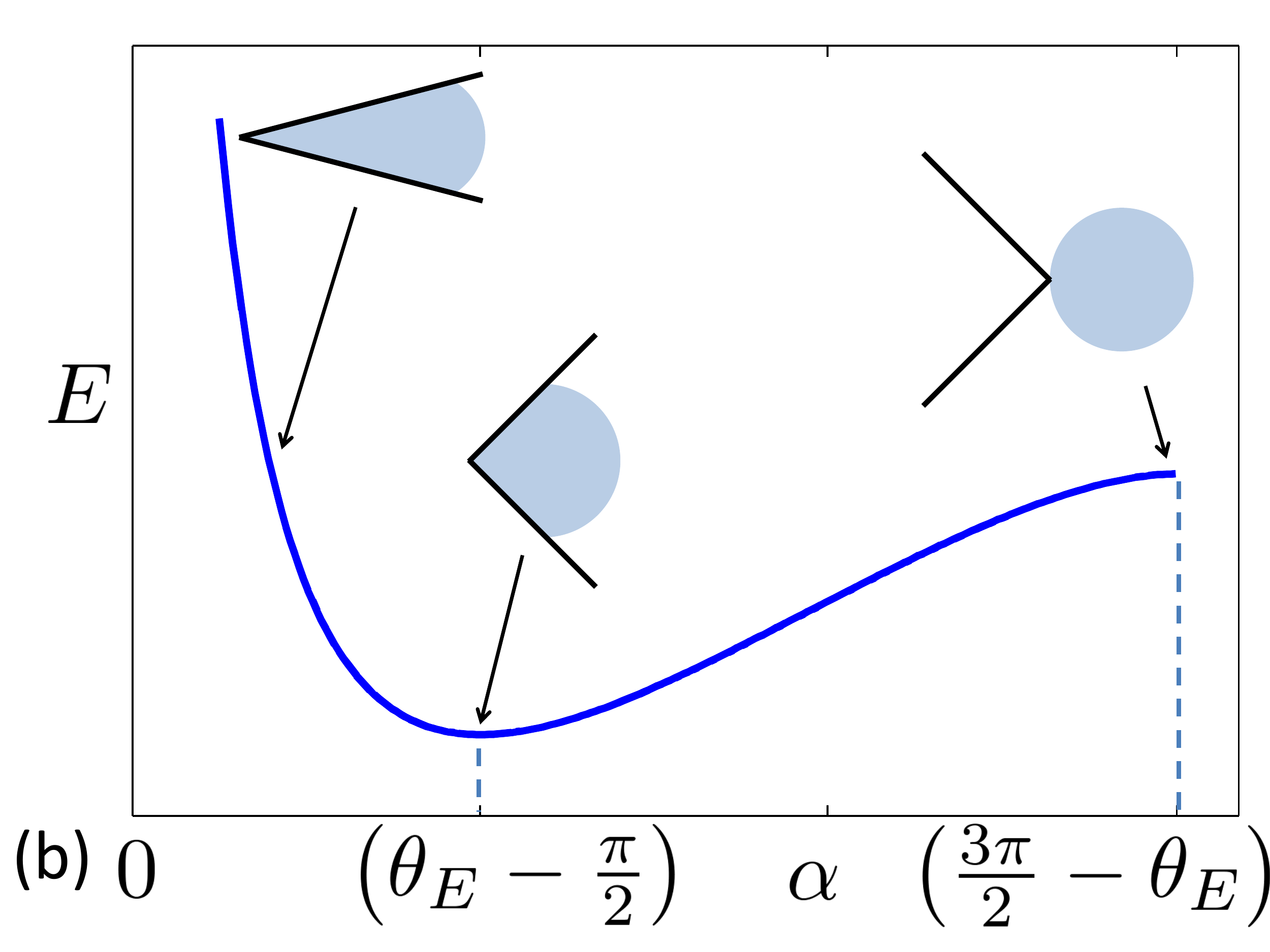}
\end{minipage}
\caption{Total surface energy, $E$,  including  solid-solid, solid-liquid, and liquid-gas surfaces (arbitrary units) as a function of the folding angle, $\alpha$ (radians), {at constant volume}. (a):  Hydrophilic case ($\theta_{E} = {\pi}/{6}$); The minimum energy is obtained for complete folding. (b): Hydrophobic case ($\theta_{E}=3\pi/4$);  The stable folding angle is given by $\alpha=\theta_{E}-{\pi}/{2}$, which is $\pi/4$ here.}
\label{fig:theta_petit}
\end{figure}

Let us first consider a two-dimensional system composed of two rigid, freely articulated infinite walls (Fig.~\ref{fig:coin_2D}).  The fluid droplet  is located at the intersection of the two  walls which are opened by a semi-angle $\alpha$ termed the folding angle. The equilibrium contact  angle  of the droplet on the solid is denoted $\theta_{E}$. A direct study of the total surface energy of the system (Fig.~\ref{fig:theta_petit}) shows that the wetting characteristics of the droplet on the wall (hydrophobic vs.~hydrophilic) plays an important role in determining the folding angle \footnote{Due to the absence of  intrinsic length scales to the problem, the   surface energies in Fig.~\ref{fig:theta_petit} are plotted with arbitrary units. Consequently,  only  the relative, instead of absolute, energy values are important.}. 
 Using the subscripts ``SL'', ``AL'' and ``SA'' to refer to the solid-liquid, air-liquid and solid-air interfaces respectively, and using $\gamma$ to denote the surface energies, 
Young's equation is given by $\gamma_{SL}+\gamma_{AL} \cos(\theta_E) = \gamma_{SA}$ where  $\theta_{E}$ is the equilibrium contact angle of the droplet on the surface.  The hydrophobic case is the one in which   $\gamma_{SL}>\gamma_{SA}$ and  $\theta_{E} > {\pi}/{2}$. Similarly, the hydrophilic case corresponds to  $\gamma_{SL}<\gamma_{AL}$, or  $\theta_{E} < {\pi}/{2}$.

Examining the energy profile numerically in the hydrophilic case (Fig.~\ref{fig:theta_petit}a, $\theta_E=\pi/6$), we see that it does not display a minimum until complete folding: the energy continues to monotonically decrease with the folding angle and  tends to $-\infty$ when the opening angle tends to 0. As a result, the most energetically favorable configuration is complete folding (in this idealized geometry, the liquid-air interface is sent to infinity). 
 
 A simple scaling analysis  in the small-angle limit can be used to understand why $\alpha=0$ is the energetically-favorable configuration. Using the notation from Fig.~\ref{fig:coin_2D}, in the limit where $\alpha$ is small it is clear that we have the scaling $ L\sim \alpha R$ where $L$ is the typical size of the air-liquid interface. The total (two-dimensional) surface  energy is thus given by $E\sim 2 R (\gamma_{SL}-\gamma_{SA}) + \alpha R \gamma_{AL}$ which, using Young's formula, gives the scaling 
 $E\sim \gamma_{AL}(\alpha R-2R\cos\theta_E)$. The droplet volume is given by the scaling $V\sim \alpha R^2$ and thus $R$ is $R\sim(V/\alpha)^{1/2}$. We finally obtain the scaling for the surface energy given by 
 \begin{equation}\label{scaling_2D}
E\sim \gamma_{AL} V^{1/2} (\alpha^{1/2}-2\alpha^{-1/2}\cos\theta_E).
\end{equation}
Clearly the minimum of Eq.~\eqref{scaling_2D} is obtained for $\alpha=0$, which is complete folding.

\begin{figure}[t]
\includegraphics[width=.6\textwidth]{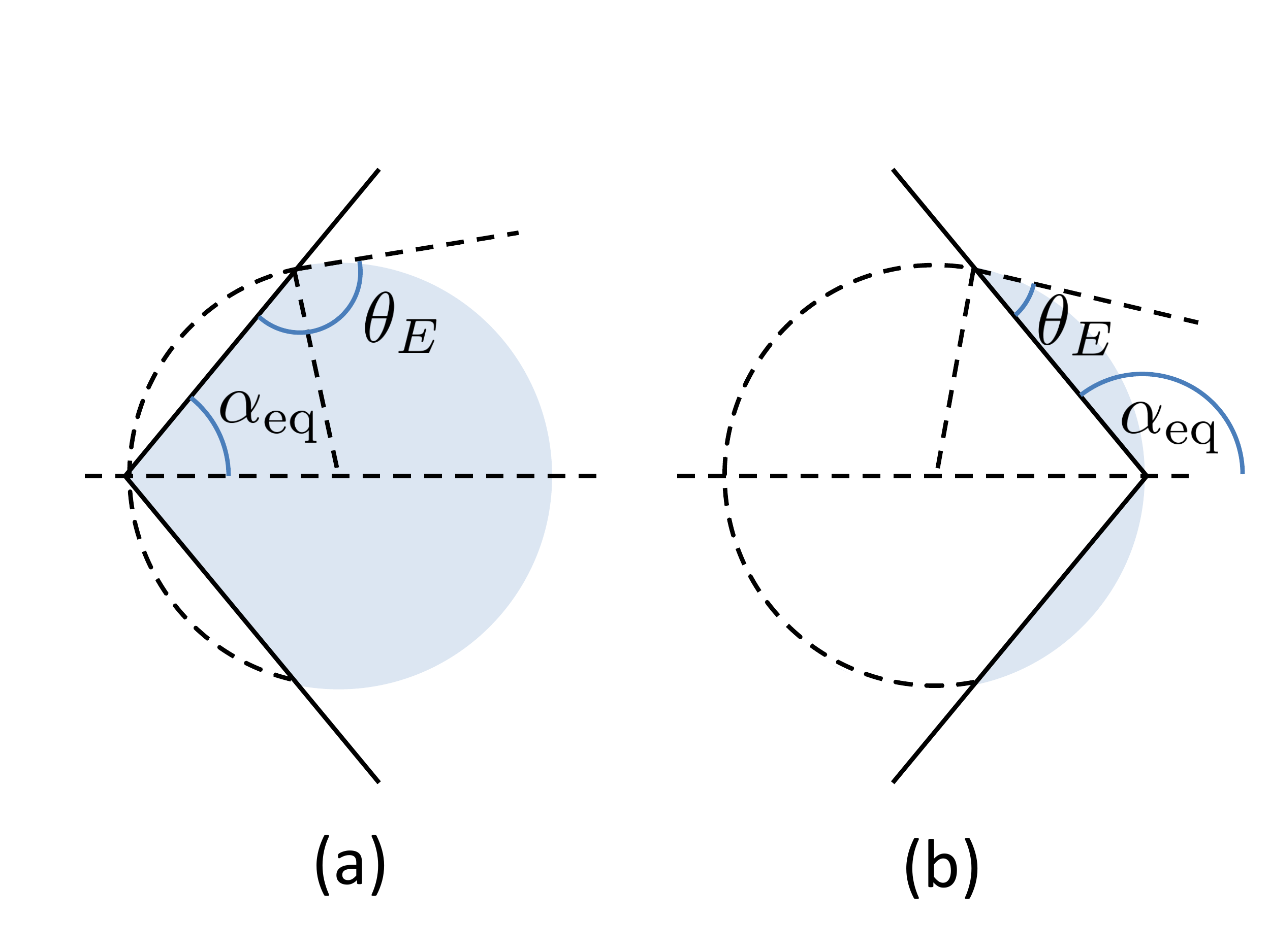}
\caption{(a): Hydrophobic folding  with $\alpha_{\text{eq}}=-{\pi}/{2}+\theta_{E}$ (energy minimum); the expansion of the droplet and the hinge intersect at the same location. (b): Hydrophilic case with $\alpha_{\text{eq}}={\pi}/{2}+\theta_{E}$ (energy maximum);  here also, the droplet and the two walls intersect at the same point.}
\label{fig:configurations}
\end{figure}

As a difference with the hydrophilic situation,  the hydrophobic regime leads to   a stable minimum of energy at a finite value of the folding angle.  This is illustrated in Fig.~\ref{fig:theta_petit}b for $\theta_E=3\pi/4$. This equilibrium opening angle can be  derived by a direct approach considering the forces acting on the surface. In order to determine the relationship between the contact angle, $\theta_{E}$ and the stable folding angle, $\alpha_{\text{eq}} $, a simple balance of forces between the contribution of the contact line  in the direction normal  to the surface and the capillary pressure leads to  the relationship
\begin{equation}\label{ideal}
\alpha_{\text{eq}} = -\frac{\pi}{2} + \theta_{E}.
\end{equation}

To derive Eq.~\eqref{ideal}, consider the setup in Fig.~\ref{fig:coin_2D}. The capillary pressure inside the droplet subjects the wall to the moment $R^2 \Delta p 2^{-1}=R^2\gamma (2r)^{-1}$ with respect to the hinge, while the direct surface tension force induces the moment $\gamma R \sin(\theta_E)$. The balance of moment then results in 
\begin{equation}
\frac{R}{r} = 2\sin(\theta_E)\cdot
\end{equation}
Noting that the ratio $R/r$ can also be written
\begin{equation}
\frac{R}{r} = \frac{\sin(\alpha+\theta_E-\frac{\pi}{2})}{\sin(\alpha)},
\end{equation}
we then obtain
\begin{equation}
\cos(\alpha-\theta_E) = 0
\end{equation}
whose solution is $\alpha_\text{eq} = -{\pi}/{2} + \theta_{E}$ or $\alpha_\text{eq} = {\pi}/{2} + \theta_{E}$ depending on the case.

\begin{figure}[t]
\includegraphics[width=.55\textwidth]{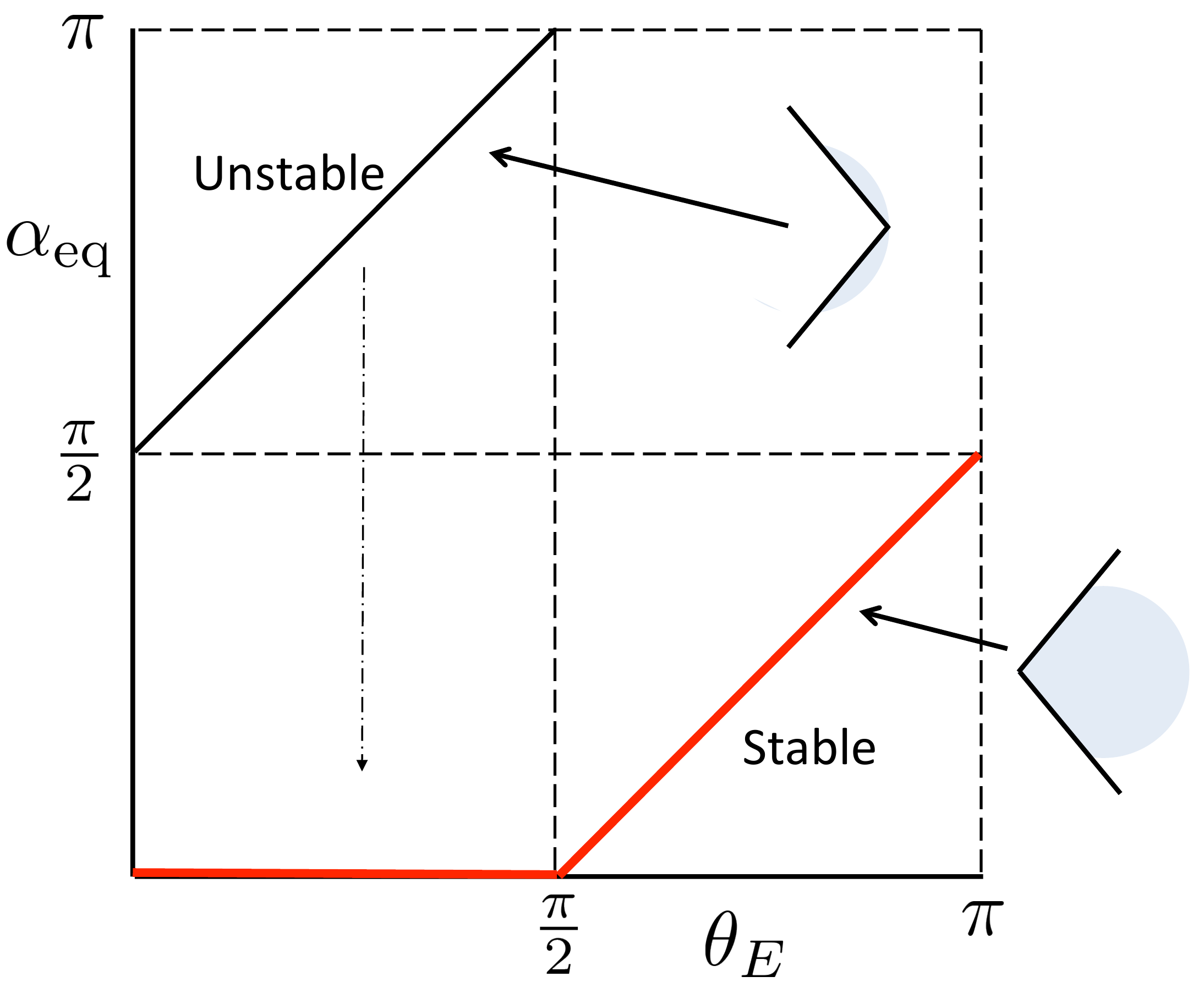}
\caption{Equilibrium folding angle, $\alpha_\text{eq}$, as a function of the contact angle, $\theta_E$. In the hydrophilic case ($\theta_{E}<\pi/2$), the system folds completely. In contrast, in the hydrophobic case,  $\theta_{E}>\pi/2$, the folding angle is finite and follows the relationship $\alpha_\text{eq}=-{\pi}/{2}+\theta_{E}$.}
\label{fig:recapitulatif}
\end{figure}

The geometry of folding is thus fully determined by the wetting property of the liquid droplet on the solid surface.  Notice that the maximum possible folding angle in this situation is predicted to be $\pi/2$, meaning that the two walls become aligned with each other (open configuration). From a geometrical standpoint, the relation obtained in Eq.~\eqref{ideal} actually corresponds to a specific configuration in which the expansion of the circular shape of the droplet exactly intersects the hinge point (see Fig.~\ref{fig:configurations}a). A similar geometry arises if we look for a balance of forces in the hydrophilic case  (see Fig.~\ref{fig:configurations}b). 
 We do obtain a metastable equilibrium with the relation $\alpha_{\text{eq}} = {\pi}/{2} + \theta_{E}$,  corresponding to the limiting case where the droplet and the walls intersect at the same point.  The folding  results are summarized in Fig.~\ref{fig:recapitulatif}. Importantly, by varying the droplet contact angle {in the hydrophobic region,} all possible folding geometries can be obtained, from completely open ($\alpha_{\text{eq}}=\pi/2$) to completely closed ($\alpha_{\text{eq}}=0$).

\subsection{Finite-size folding}
\label{finite-size}
We next examine the situation where the walls have finite size. For simplicity we assume that both sides have the same length. In that case, in addition to the contact angle, the droplet volume  will impact  the folding angle. Let us consider a fixed equilibrium contact angle, $\theta_{E}$. In the hydrophobic case, for a sufficiently small volume of liquid located near the hinge, the problem is just the same as in the previous section (in other words the walls appear to be infinite). Therefore, under a certain critical volume, the folding angle obeys the relation previously derived, $\alpha_{\text{eq}} = -{\pi}/{2} + \theta_{E}$. The critical volume corresponds to the point where the interface of the droplet reaches the edges of the walls. That critical volume depends thus  on both the wall length  and the equilibrium contact angle and can be easily computed (alternatively, one can compute the critical folding angle). If the droplet volume is further increased beyond this critical value, the contact line remains pinned at the edge of the walls (equivalently, the contact area remains fixed).

\begin{figure}[t]
\includegraphics[width=.55\textwidth]{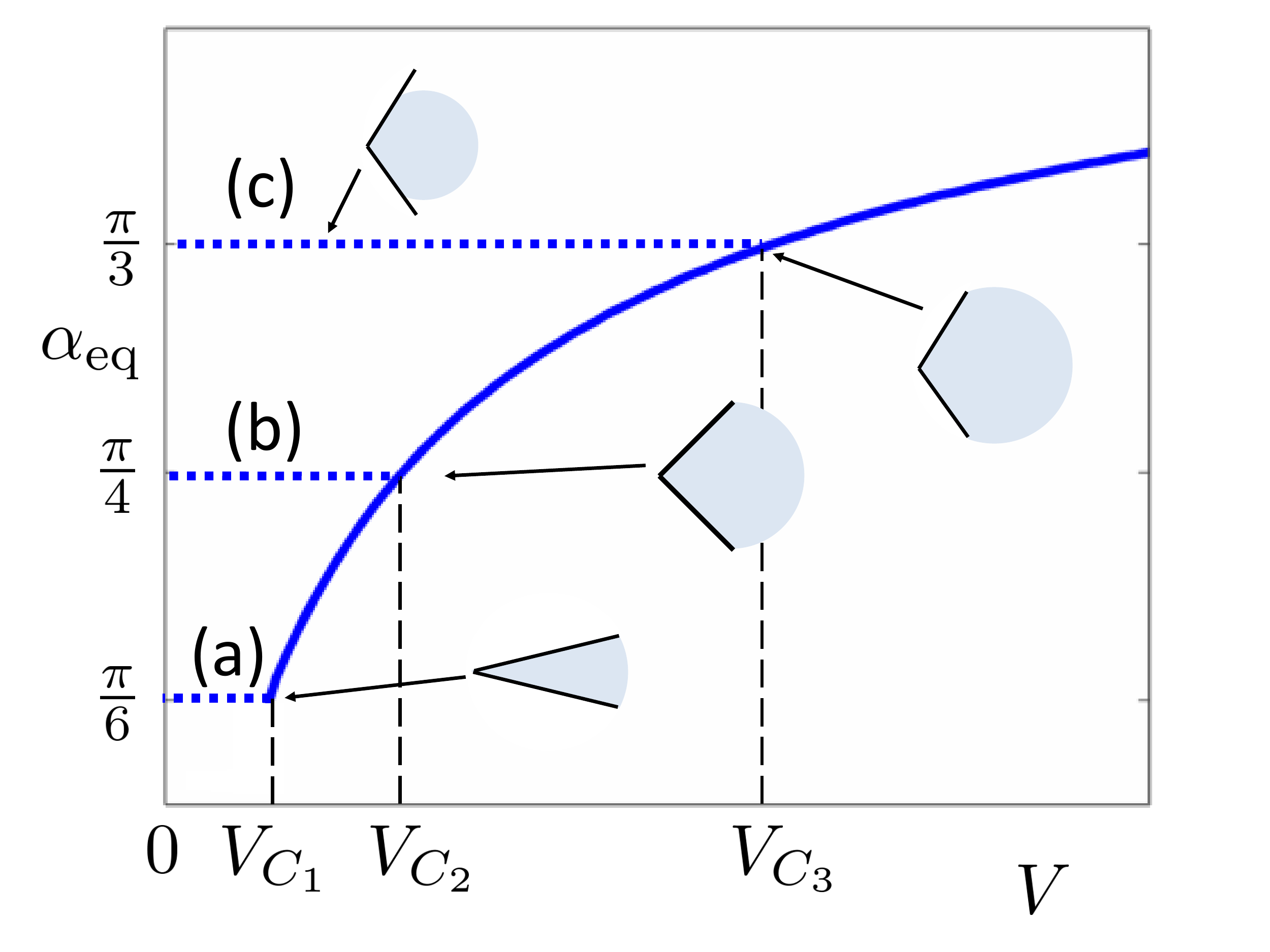}
\caption{Folding angle, $\alpha_\text{eq}$ (radians), as a function of the droplet volume,   $V$ (arbitrary units), for finite-size walls.  
Dotted line: before the critical volume, the droplet does not reach the edges of the walls and the opening angle remains constant. {The cases depicted here correspond to (a) $\theta_E=2\pi/3$, (b) $\theta_E=3\pi/4$, and (c) $\theta_E = 5 \pi/6$}. Continuous lines: after the critical volume, the contact line is pinned and the opening angle increases monotonically. The critical volume depends on the value of contact angle (or, alternatively, on the critical folding angle, as shown here).}
\label{fig:pb_au_bord}
\end{figure}

Two different methods can be applied to study this situation. As a first option, we can investigate the energy profile numerically for a given fixed volume as a function of the folding angle  (or the contact angle), and verify that there is still a minimum of energy at a given folding angle. Alternatively, we can also notice that the relationship $\alpha_{\text{eq}} = -{\pi}/{2} + \theta$ still holds in that case, due to the balance of forces perpendicular to each wall, provided $\theta>\theta_{E}$ is understood as  the apparent contact angle.  The same geometrical feature as seen previously (see Fig.~\ref{fig:configurations}) applies here. In particular,  the continuation of the circular  droplet exactly intersects the hinge of the walls, making the calculation of the volume for a given folding angle $\alpha$ straightforward, and we obtain
\begin{equation}\label{aftercritical}
V(\alpha) = \frac{L^{2}}{2}\tan \alpha+\frac{\alpha L^{2}}{2 \cos^{2}\alpha },
\end{equation}
for $0\leq \alpha < \pi/2$. 

In summary, for a fixed contact angle, the folding angle remains constant and given by Eq.~\eqref{ideal} below the critical droplet volume, after which the  volume-folding angle relationship is given by Eq.~\eqref{aftercritical}. 
The results are illustrated  numerically in Fig.~\ref{fig:pb_au_bord} in the cases $\theta_E=2 \pi/3$, $\theta_E= 3\pi/4$, and $\theta_E = 5\pi/6$.  Note that the  folding angle converges to $\pi/2$ in the limit of large droplet volumes, which can also be seen by evaluating Eq.~\eqref{aftercritical} in the limit $V\to \infty$.

\subsection{Folding of curved walls}
\label{curved}

\begin{figure}[t]
\includegraphics[width=.65\textwidth]{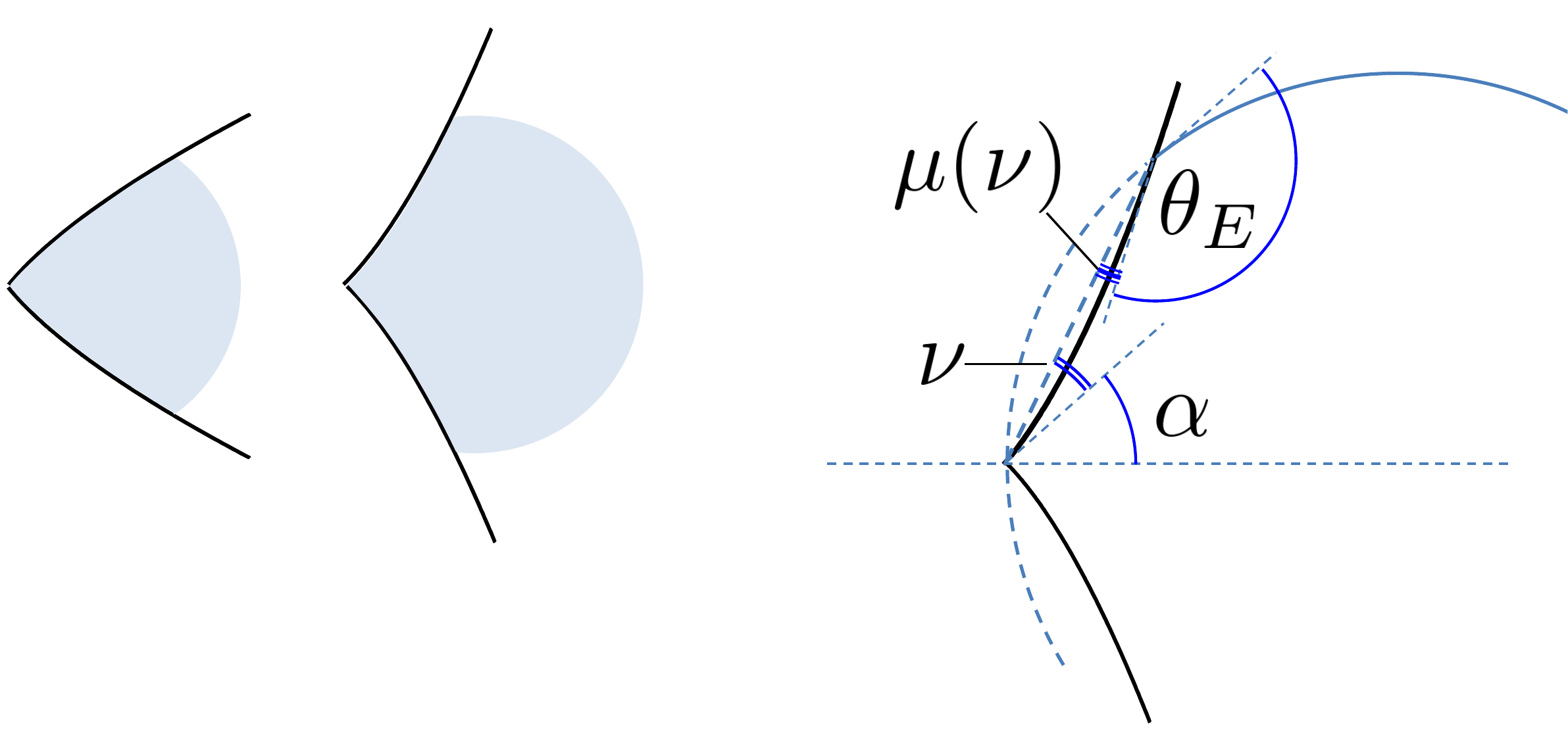}
\caption{Left: Shapes chosen for concave and convex walls. 
Right: For a given $\alpha$  and a known wall shape, we can calculate the corresponding angles $\nu$ and $\mu(\nu)$. Mechanical equilibrium requires that the expansion of the arc of circle of the droplet intersects the hinge of the walls, which is equivalent to the relationship $\alpha + \nu = -\pi/2 + \theta_{E} +\mu(\nu)$.}
\label{fig:convex-concav}
\label{fig:parametrage-convex}
\end{figure}

We now consider  another potential feature to the system, namely the fact that the walls might not be perfectly straight. Here we focus on smooth curved surfaces, and address both convex and concave walls. For illustrative purposes, we consider the  particular case of walls which shapes are given by the equation $y = x + x^2$ (convex) and its reciprocal function $y = -1/2+\sqrt{1/4+x}$ (concave), as illustrated in Fig.~\ref{fig:convex-concav} (left). In both cases, the equation refers to the shape of the topmost wall when $\alpha=\pi/4$, and the bottom wall is assumed to be its mirror image. We define the folding angle as the semi-angle between the tangents of the two walls at the hinge. 

\begin{figure}[t]
\includegraphics[width=.5\textwidth]{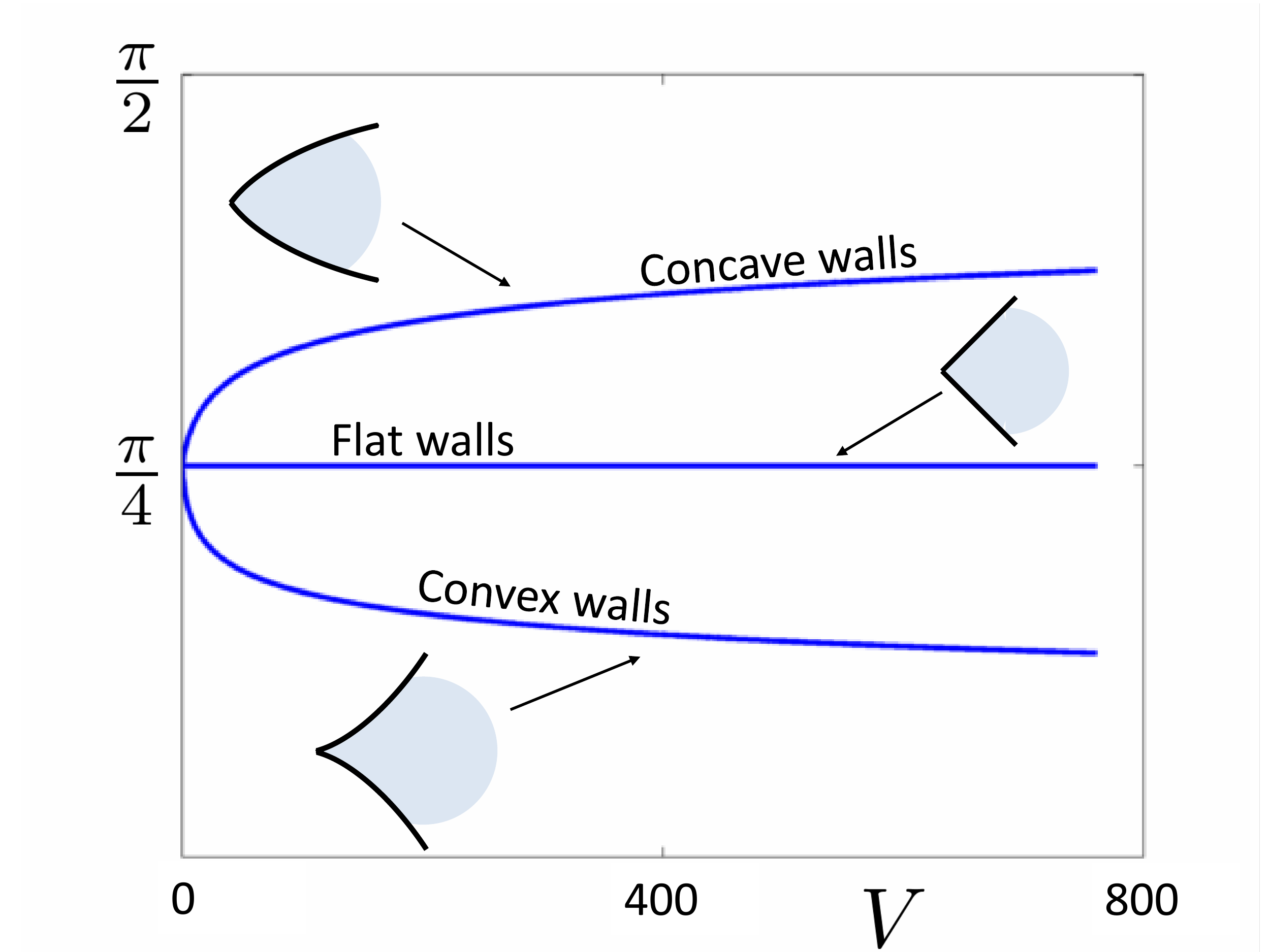}
\caption{Equilibrium folding angles, $\alpha_{\text{eq}}$, for concave (top), flat (middle) and convex (bottom) boundaries, for a contact angle $\theta=3\pi/4$, as a function of the droplet volume, $V$. The volume is given in normalized units corresponding to the two curves $y=x+x^2$ and the reciprocal function given by $y=({-1+\sqrt{1+4x}})/{2}$. For concave walls  the folding angle $\alpha$ increases with the volume $V$ while it decreases in the  convex case. Note that the two curves are not symmetric from each other.}
\label{fig:convex-concav-3}
\end{figure}

As in the previous section,  two different approaches can be exploited to derive the  relationship between folding angle and droplet  volume. We can elect to compute  directly the surface energies for different folding angles at a given droplet volume and determine  the angle leading to the smallest energy. Alternatively,  we can use the local balance of forces as illustrated in  Fig.~\ref{fig:parametrage-convex} (right). The relationship becomes $\alpha + \nu = -\pi / 2 + \theta_{E} +\mu(\nu)$, with the angle $\nu$ and $\mu(\nu)$ defined geometrically in Fig.~\ref{fig:parametrage-convex} (right) to find the corresponding flat-wall problem. That relationship can be exploited to  plot the folding angle,  $\alpha_\text{eq}$, as a function of the droplet volume, with results shown in Fig.~\ref{fig:convex-concav-3}. In the case of convex walls (resp.~concave walls), the folding angle decreases (resp.~increases) with an increase in the droplet volume.

\subsection{Folding of walls with discontinuous slopes (kinks)}
\label{kinks}

\begin{figure}[t]
\includegraphics[width=.65\textwidth]{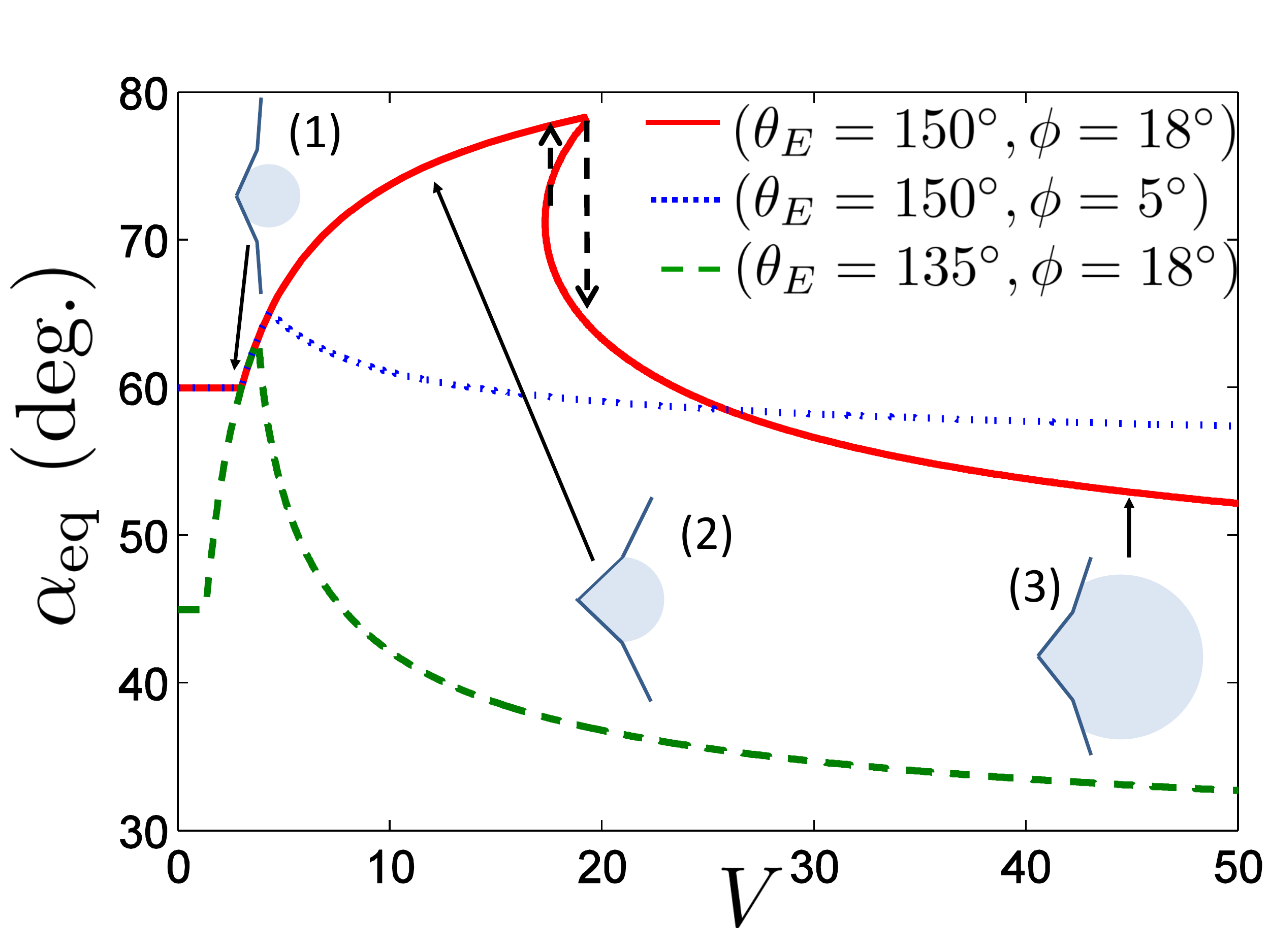}
\caption{Equilibrium folding angle, $\alpha_{\text{eq}}$, as a function of  the droplet volume, $V$, when there is a kink along the walls ($\theta_{E}=5 \pi/6$).  As the droplet volume increases, three regimes appear: (1) The folding angle $\alpha_{\text{eq}}$ is  constant and  the configuration  similar to the infinite-sized wall; (2): When the interface reaches the kink, the contact line remains pinned, as in the finite-size folding problem, and the folding angle increases; (3): When the angle of contact with the second part of the wall reaches $\theta_{E}$, the triple line moves again, and $\alpha_{\text{eq}}$ decreases, possibly with a sudden jump (hysteresis).}
\label{fig:kink_result2}
\end{figure}

\begin{figure}[t]
\includegraphics[width=.45\textwidth]{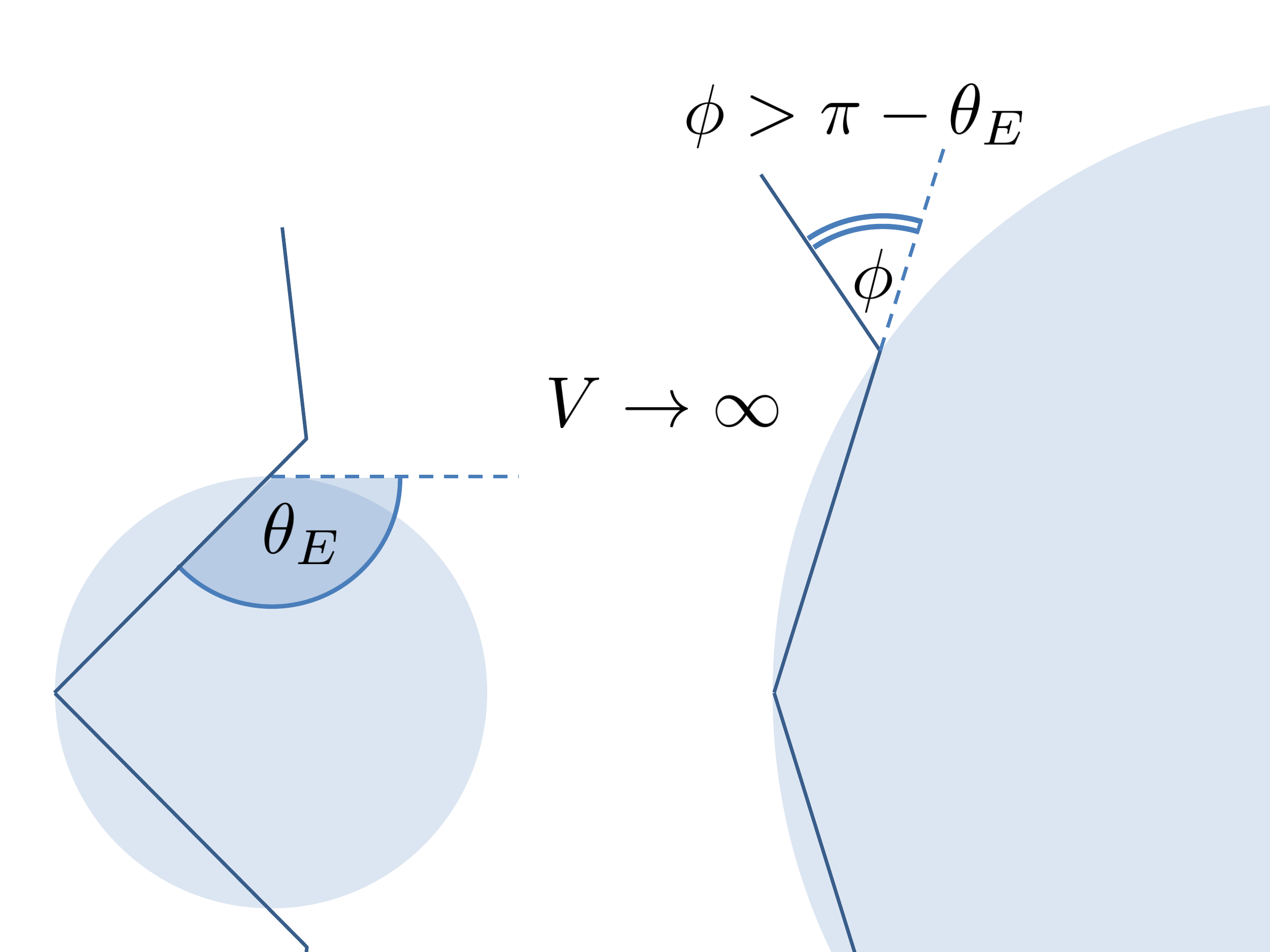}
\caption{If the kink angle, $\phi$, is larger than $\pi-\theta_E$, the contact line remains pinned at the kink and the  system behave similarly to the finite folding case studied in \S II.B}
\label{fig:2regimes}
\end{figure}

Instead of a continuous change in wall shape, the geometry of the surface could undergo a discontinuous slope change and be kinked.  This is the situation we address here. We consider a wall  described by $y=x$ ($y=-x$ for the symmetric wall) for $0\leq x \leq 1$ and $y=\tan ( \phi+\pi/4)x-1$ ($y=-[\tan(\phi+\pi/4)x-1]$ for its symmetric wall) for $x\geq1$, where the angle $\phi$ is defined in Fig~\ref{fig:2regimes}.  Following the same method as above, we obtain the numerical results  shown in Fig~\ref{fig:kink_result2} for several values of $\theta_E$ and $\phi$, and featuring three different regimes.
The first two portions of the curve correspond to the finite-size folding problem. Below a critical droplet volume, the folding angle  follows  $\alpha_{\text{eq}} = - \pi/2 + \theta_{E}$, similarly to the  infinite-wall situation. Then the droplet reaches the kink, and the contact line gets pinned, giving a problem equivalent to the finite-size wall folding problem. This is valid until the volume reaches a new limit, corresponding to the exact moment when the apparent angle between the droplet interface and the second part of the wall equals the actual contact angle $\theta_{E}$. In the  particular case considered in the figure,  $(\theta_E = 150^\circ,\phi=18^\circ)$,  we  see that this limit is accompanied by a sudden jump in the folding angle, and  hysteresis. Such feature is absent when the angle between the two walls, measured by $\phi$, is small or if the contact angle is small enough. Finally, if the angle $\phi$ at the kink is larger than $\pi-\theta_{E}$,  the contact angle can never reach $\theta_{E}$ (see  Fig.~\ref{fig:2regimes}) and the relationship between volume and folding angle is  equivalent to the finite-wall problem.

\subsection{Two-droplet folding}
\label{2drop}

The situations considered so far addressed the influence of contact angle, droplet volume, and wall shape on the folding by surface tension. We focus here on the case in which fluid is present on both sides of the hinge;  in two-dimensions this is equivalent to saying that we have two droplets. Assuming the walls to be of infinite extent, the problem with two droplets is characterized by three dimensionless parameters: the two equilibrium contact angles, denoted here $\theta_1$ and $\theta_2$, and the ratio of droplet volumes, $V_1/V_2$, assumed to be above 1 without loss of generality.

\subsubsection{Identical contact angles}

\begin{figure}[t]
\includegraphics[width=.35\textwidth]{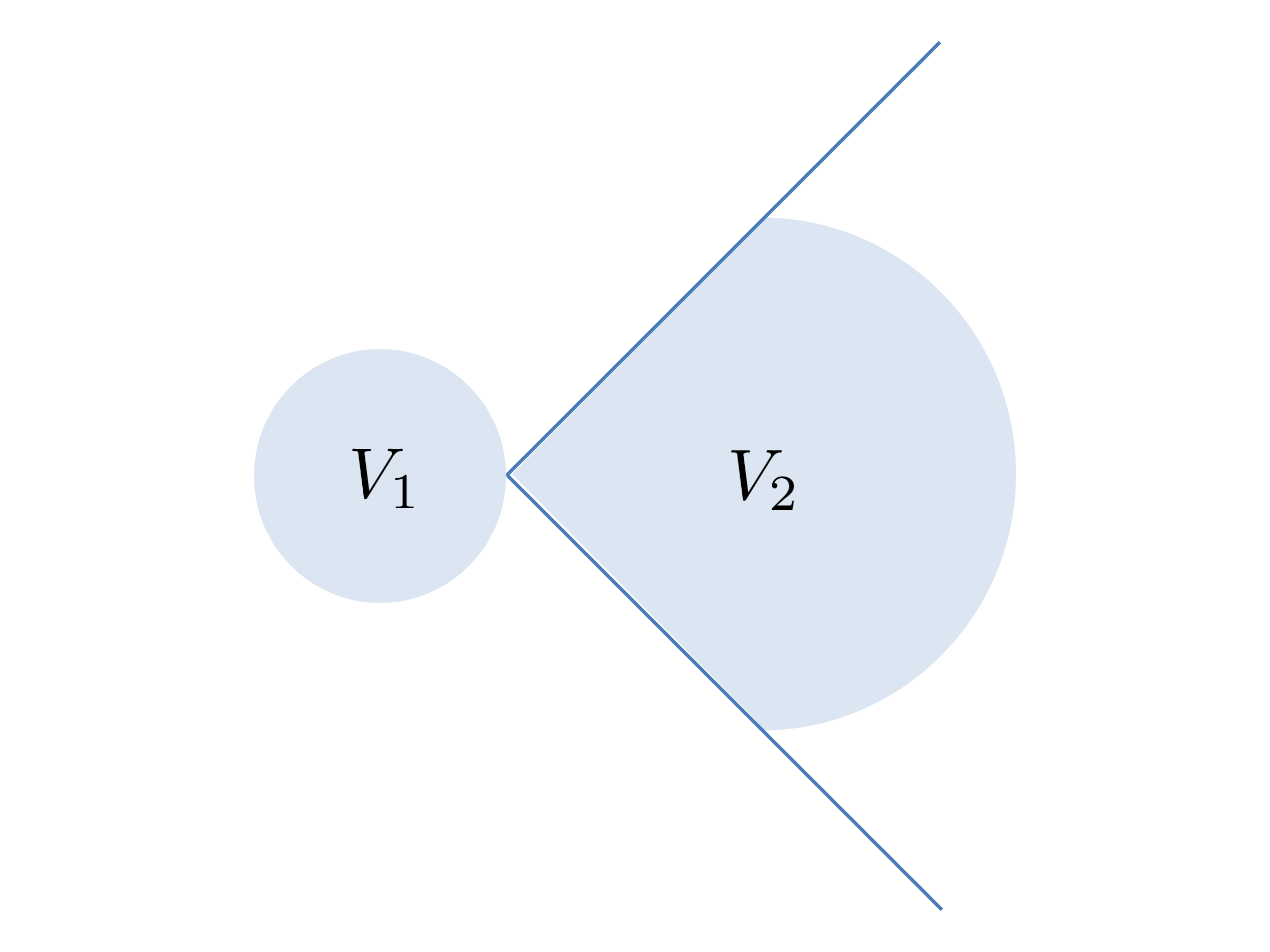}
\caption{In the situation with two droplets from the same liquid, the stable equilibrium configuration has one droplet satisfying the single-droplet equilibrium,  $\alpha_{\text{eq}}=\theta_{E}-{\pi}/{2}$, while the  other one merely touches the hinged point.}
\label{fig:two_drops_0}
\end{figure}

Both the energetic approach and the one based on the balance of forces show that there exists a constant $C>1$ which depends on 
$\theta_1=\theta_2\equiv \theta_{E}$ so that if $V_{1}/V_{2} > C$, i.e if  droplet 1 is sufficiently larger than droplet 2, there is {only} one stable equilibrium position, in which droplet 1 behaves as if isolated, $\alpha_\text{eq} = -{\pi}/{2} + \theta_{E}$, while droplet 2 touches the walls only at the hinged point (see Fig.~\ref{fig:two_drops_0}). In contrast, if $ V_{1}/V_{2} < C$, {both} configurations mentioned above correspond to a stable equilibrium. Between these two positions, there exists a unstable equilibrium (i.e. a maximum of energy).

\subsubsection{Different contact angles}
\begin{figure}[b]
\includegraphics[width=.65\textwidth]{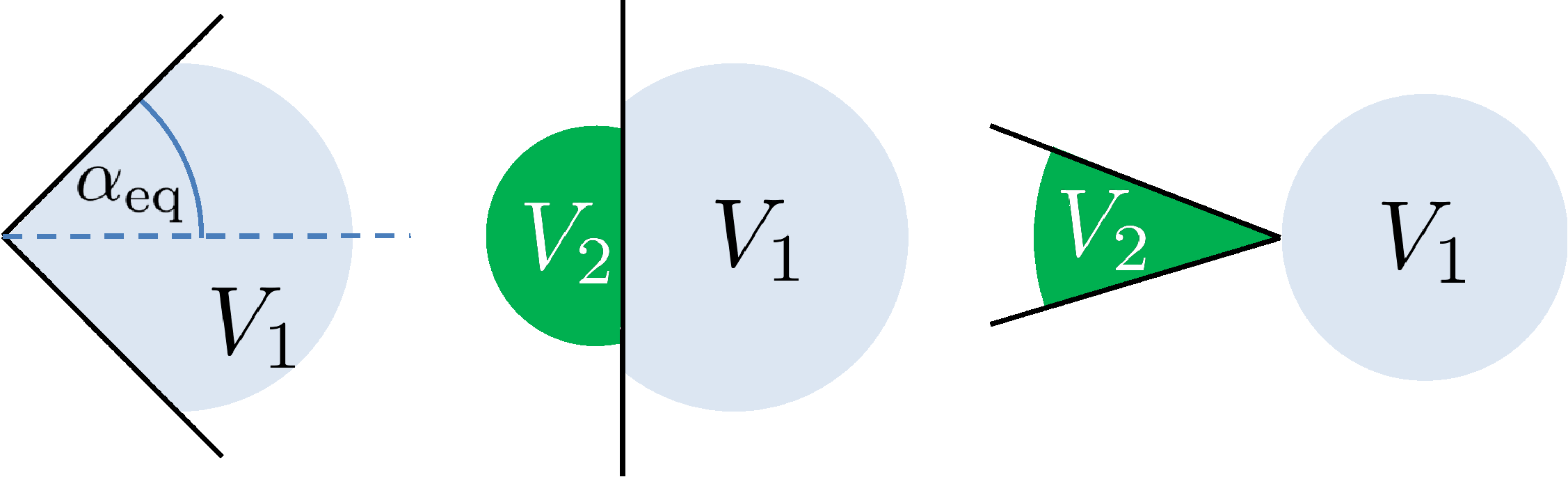}
\caption{Variation of the folding angle with the volume ratio if the additional droplet (\#2) is more hydrophilic than the the original droplet (\#1).}
\label{fig:two_drops}
\end{figure}

\begin{figure}[t]
\includegraphics[width=.55\textwidth]{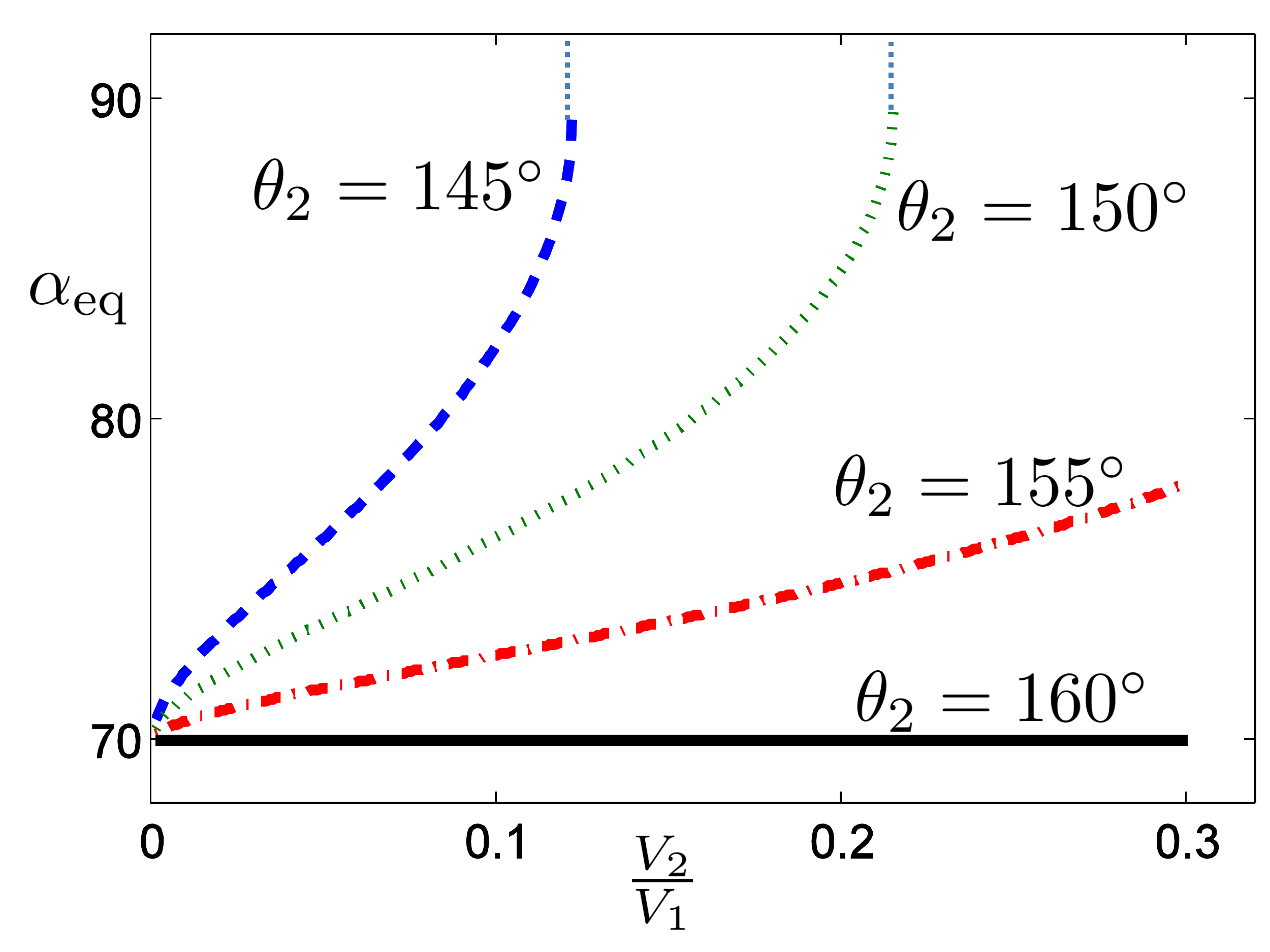}
\caption{Folding with two hydrophobic droplets: Equilibrium folding angle as a function of the ratio of volumes.  Droplet \#1 has a contact angle $\theta_{1}=160^\circ$ and we consider three different contact angles for droplet \#2 (145$^\circ$, 150$^\circ$, and 155$^\circ$). Without  droplet \#2, the equilibrium folding angle  would be  $\alpha_{\text{eq}}=\theta_{1}-90^\circ=70^\circ$). The most hydrophilic droplet pulls the system closer towards its own equilibrium folding angle. 
When the folding angle reaches $\pi/2$ (open configuration), the instability reverses the folding into the situation of Fig.~\ref{fig:two_drops} (right) with droplet \#2 inside and  \#1 outside.}
\label{fig:lieux_deux_gouttes}
\end{figure}

Let us consider the case in which we start with a single hydrophobic droplet (\#1) in the equilibrium folded configuration. We then add a second hydrophobic droplet on the other side of the hinge (\#2) and progressively increase its size (see Fig.~\ref{fig:two_drops}). If droplet \#2 is more hydrophobic than \#1, the only stable equilibrium is the same as the one depicted in Fig.~\ref{fig:two_drops_0}). In contrast, if \#2 is more hydrophilic, then surface energies from the two droplets lead to a modification of the equilibrium folding. An increase of $V_2/V_1$  leads to an increase of the folding angle until a critical angle of $\pi/2$, after which the folding angle reverses and is such that $\pi-\alpha$ is equal to the equilibrium folding angle of droplet \#2 while droplet \#1 barely touches the hinge, as in Fig.~\ref{fig:two_drops_0}.  This is illustrated numerically in Fig.~\ref{fig:lieux_deux_gouttes} in the case where $\theta_{1}=160^\circ$ for three different wetting angles for droplet \#2.

\subsection{De-wetting from the hinge}
\label{dewet}
In all  cases analyzed in the previous sections, we have  assumed that the droplet was present at the hinge point all the way to the tip. Alternatively, a second free surface could appear near the hinge point whose  surface energy would need to be included.  This situation, which we refer to as ``de-wetting",   would be   of particular relevance in three dimensions where air (or the surrounding fluid of any kind) could easily reach the hinge from the sides of the droplet.

\begin{figure}[t]
\includegraphics[width=.53\textwidth]{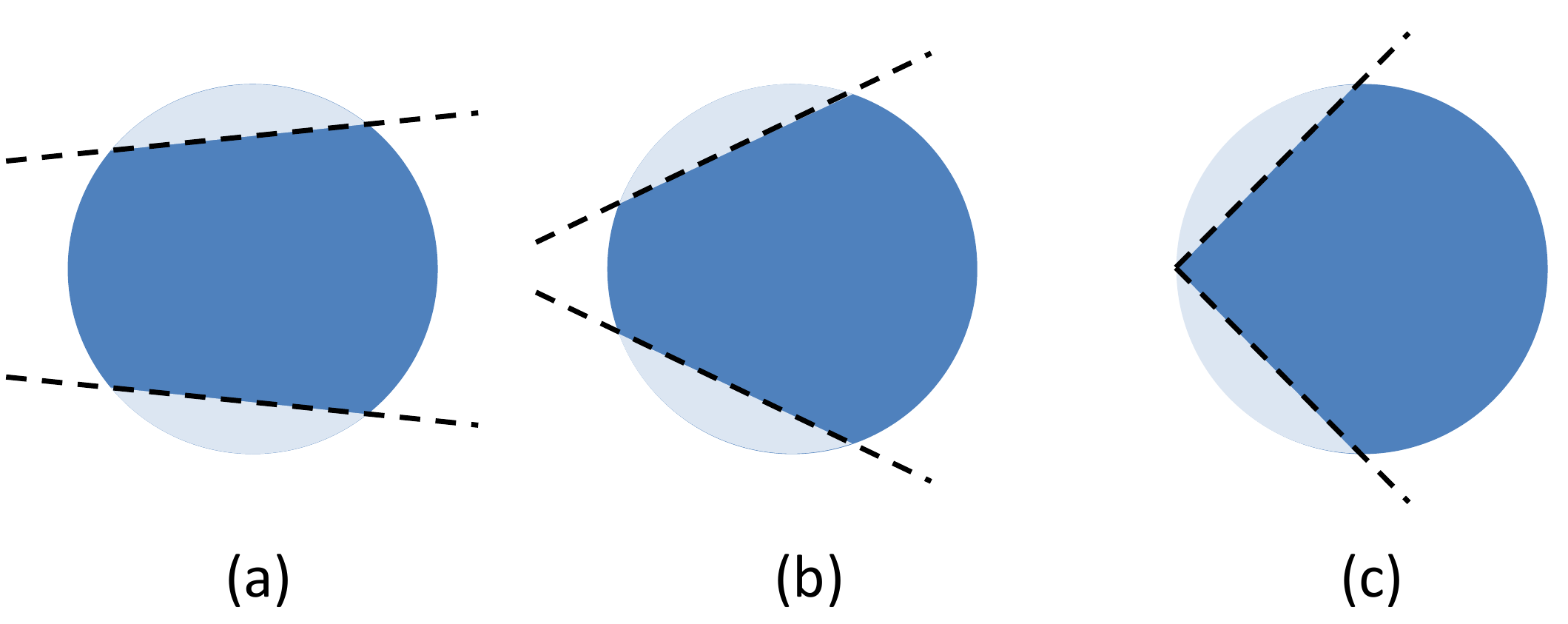}
\caption{De-wetting from the hinge leads to no change in surface energy with a change in the droplet position provided there is no contact angle hysteresis. Configurations (a), (b), and (c) have the same surface energy although the folding angles are different. In cases (a) and (b), the folding angle satisfies  $\alpha<-{\pi}/{2}+\theta_E$ while (c) corresponds to the limit case where $\alpha=\alpha_{\text{eq}}=-{\pi}/{2}+\theta_E$.}
\label{fig:dewet}
\end{figure}

\begin{figure}[t]
\includegraphics[width=.5\textwidth]{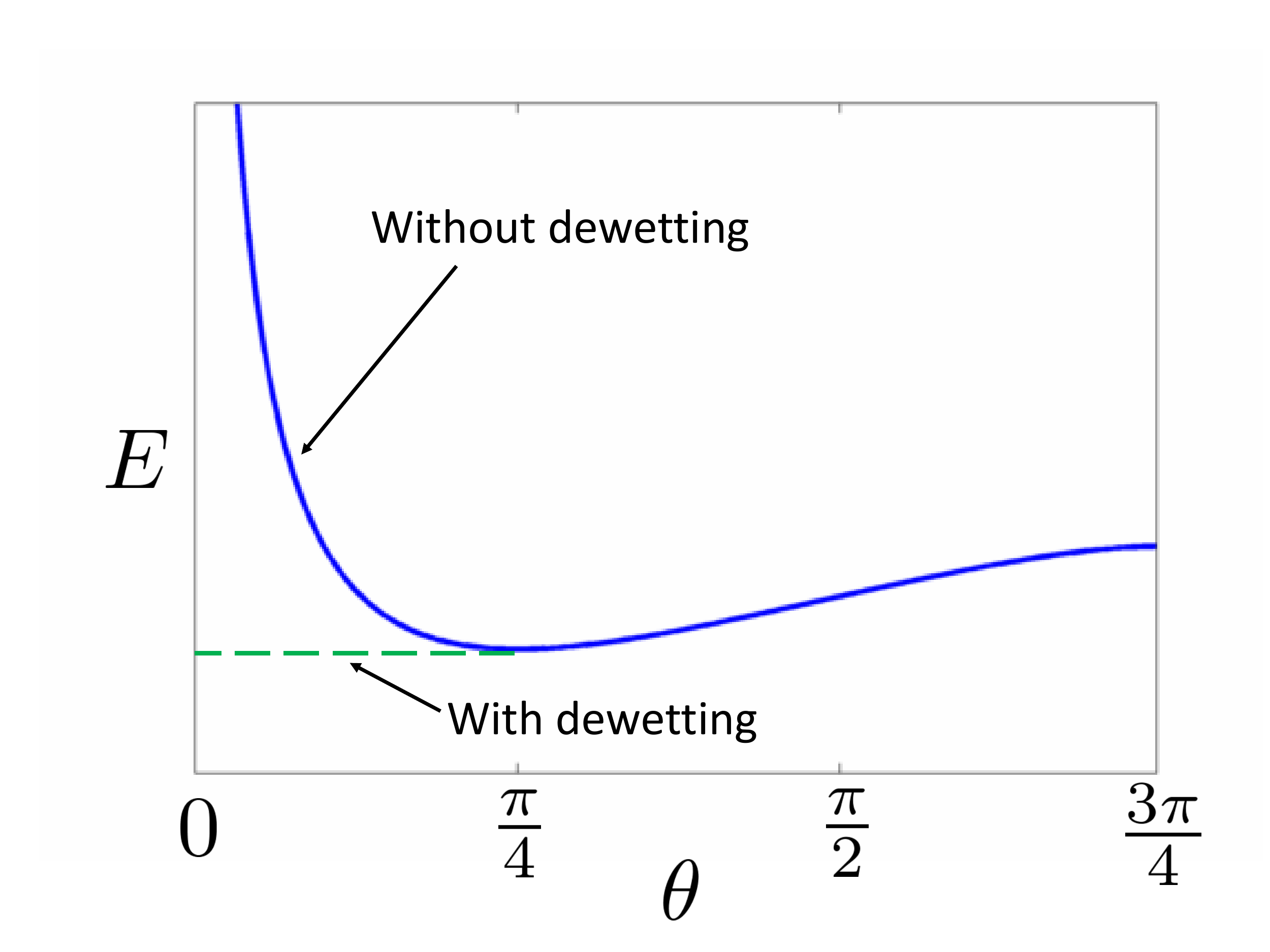}
\caption{Total surface energy $E$ (arbitrary units) as a function of the folding angle $\alpha$ for $\theta_E=3\pi/4$, in two dimensions. When we  allow de-wetting to occur, the same minimum of surface energy is obtained for the range of folding angles $0\leq \alpha\leq -\pi/2+\theta_E$.}
\label{fig:plot_dewet}
\end{figure}

Let us therefore reexamine here the model of \S\ref{single_droplet} in the case where a degree of freedom is  authorizing a second interface to be created at the hinge point. In the absence of gravity, the new interface must  have the same curvature as the external free surface  of the droplet. Because both interfaces intersect a flat surface with the same constant contact angle  they must have  the same center of curvature, which leads to the interesting geometrical property illustrated in Fig.~\ref{fig:dewet}: The droplets are both part of the same disk  from which two identical caps have been removed. The size of these caps remain identical  in order to keep a constant contact angle and droplet volume. It implies that the different areas intervening in the calculus of the energy are the same, and the energy is therefore constant. For any folding angle such that $0 \leq \alpha \leq  -{\pi}/{2}+\theta_{E}$, all configuration similar to the ones in Fig.~\ref{fig:dewet}  have the same surface energy (see Fig.~\ref{fig:plot_dewet}), and therefore all folding angle are equally energetically possible. So the presumed stable equilibrium is actually  metastable. As discussed below, this remains true in  three dimensions. However, as shown in the next section, the presence of contact angle hysteresis will in general prevent de-wetting and render the equilibrium folding angle stable.

\section{Folding with contact angle hysteresis}\label{hyst}

In the models we  considered in \S\ref{2D}, we derived the energy profile against relevant geometrical and physical parameters through direct differentiation. This was actually only made possible by the fact that $\delta E$, the variation of surface energy accompanying a small change in the parameters, was an exact differential. Real droplets behave a bit differently~\cite{degennes_book}. The triple line (at the junction air-liquid-wall) can stay fixed and, instead of moving, adopt various configurations -- a phenomenon well known as contact angle hysteresis, which can for example be due to deformations of the triple line caused by impurities and defects \cite{degennes_book}. In order to move the triple line, a net tangential force has to be applied, hence producing a work term in the variation of the energy. This is in contrast to the ideal case where the different surface tension terms always balance in the tangential direction. As a consequence, the difference of energy, $\Delta E$, between two different configurations will depend on the path followed -- in other words,  $\delta E$ is no longer an exact differential. The contact angle no longer takes a unique value but varies between a lower and an upper bound. The smallest value, denoted $\theta_{r}$, is the receding angle. It is the minimum contact angle before the triple line starts to recede in the direction of the liquid. The largest value, denoted $\theta_{a}$, is the advancing angle. It is the maximum contact angle before the triple line moves away from the fluid.

\begin{figure}[t]
\includegraphics[width=.5\textwidth]{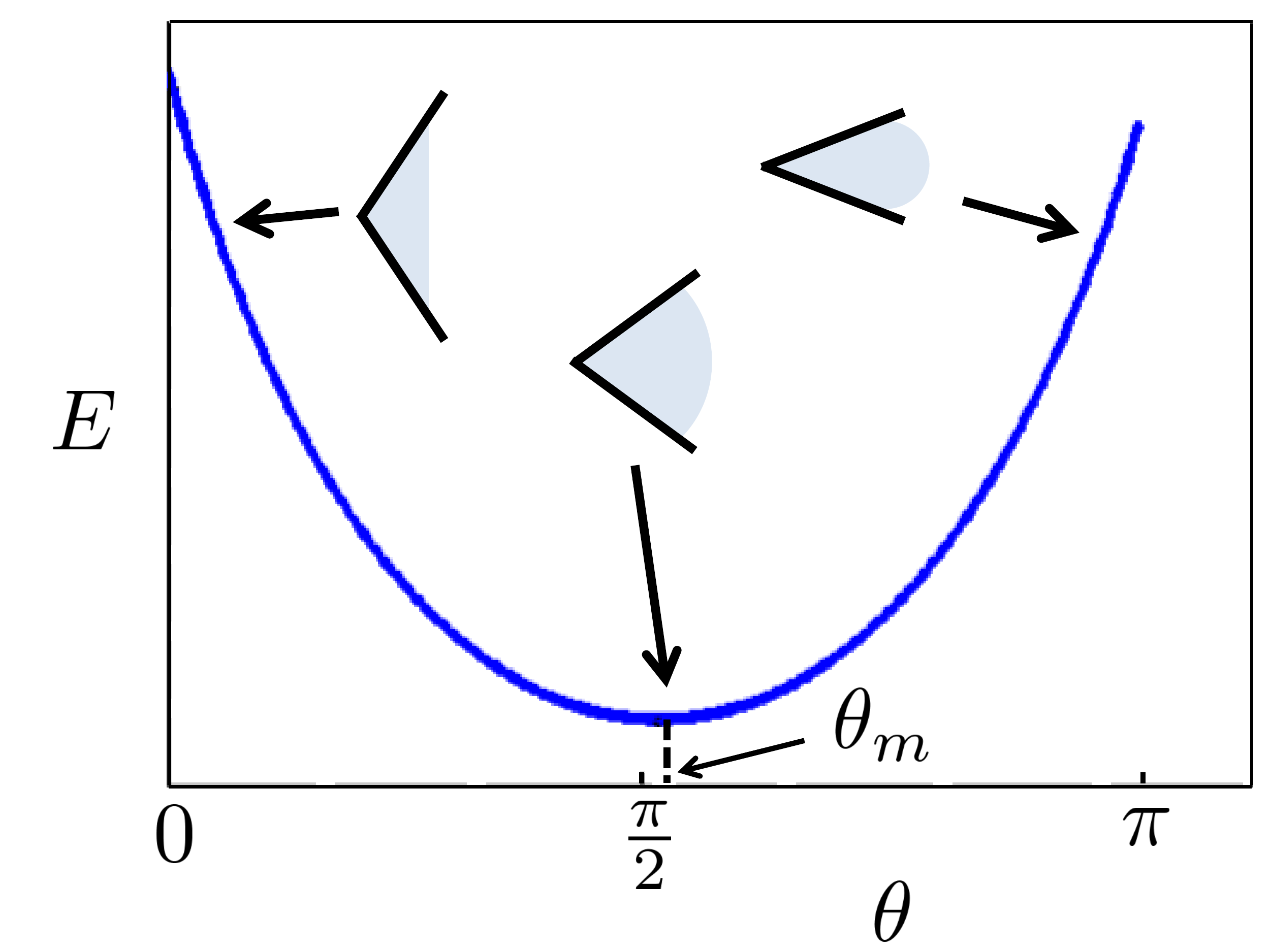}
\caption{Surface energy, $E$, as a function of the contact angle, $\theta$,  for a fixed location of the contact line. Since the volume remains constant, the folding angle $\alpha$ decreases as $\theta$ increases. The surface energy is minimal at the angle $\theta_{m}$. If $\theta_{r}\leq \theta_{m} \leq \theta_{a}$, $\theta$ and $\alpha$ will vary  until $\theta$ reaches $\theta_{m}$. 
If $\theta_{m} \leq \theta_{r}$, then $\theta$ will decrease (and $\alpha$ will increase) until it reaches $\theta_{r}$, and the contact line will recede. 
If $\theta_{a} \leq \theta_{m}$, $\theta$ will increase until it reaches $\theta_{a}$, and the contact line will advance.}
\label{fig:energy_against_theta_1}
\end{figure}

With hysteresis, the  manner in which the system gets to a local equilibrium is therefore path-dependent. The different relevant parameters are the following:  the initial non-equilibrium folding angle, the initial contact angle, the droplet volume, $\theta_{r}$, 
$\theta_{a}$,  and the various surface energies (which can be related using Young's equation and the value  of  $\theta_{E}$). As a  starting point for this problem, we first notice that as long as the contact angle stays between $\theta_{r}$ and $\theta_{a}$, the triple line does not move, so  no additional work is involved and we can study the energy directly by computing the areas of the interfaces exactly as  previously. In that case, we can either take $\alpha$ or $\theta$ as the  variable given that they are related each other by the fixed volume and fixed contact line conditions. As the inequality constraint  is on the contact angle, it is natural to use $\theta$ as the variable against which to minimize the total surface energy.  
Plotting the surface energy in Fig.~\ref{fig:energy_against_theta_1} as a function of  $\theta \in [0,\pi]$  yields a minimum of energy at $\theta = \theta_{m}$. In the general case, $\theta_{m}$ is different from $\theta_{E}$. However, it is important to note  that  the relation $\alpha = -\pi/2 + \theta_{m}$ still holds by normal force balance. 
With the value of $\theta_{m}$ determined, we compare it to both $\theta_{r}$ and $\theta_{a}$ and  distinguish between three different cases.

\subsection{Case $\theta_{r}\leq \theta_{m} \leq \theta_{a}$}

  If $\theta_{r}\leq \theta_{m} \leq \theta_{a}$, and provided the initial configuration also lies within this range, the system will directly find its minimum of energy without requiring a motion of  the triple line. The final folding angle will then follow  $\alpha=-{\pi}/{2}+\theta_{m}$.  If $\theta_{m}$ is outside the range $[\theta_{r},\theta_{a}]$, then the contact line will necessarily move before reaching a local equilibrium, with two different situations to distinguish: $\theta_{m} < \theta_{r}$ and $\theta_{m} > \theta_{a}$.

\subsection{$\theta_{m} < \theta_{r}$}

\begin{figure}[t]
\includegraphics[width=.5\textwidth]{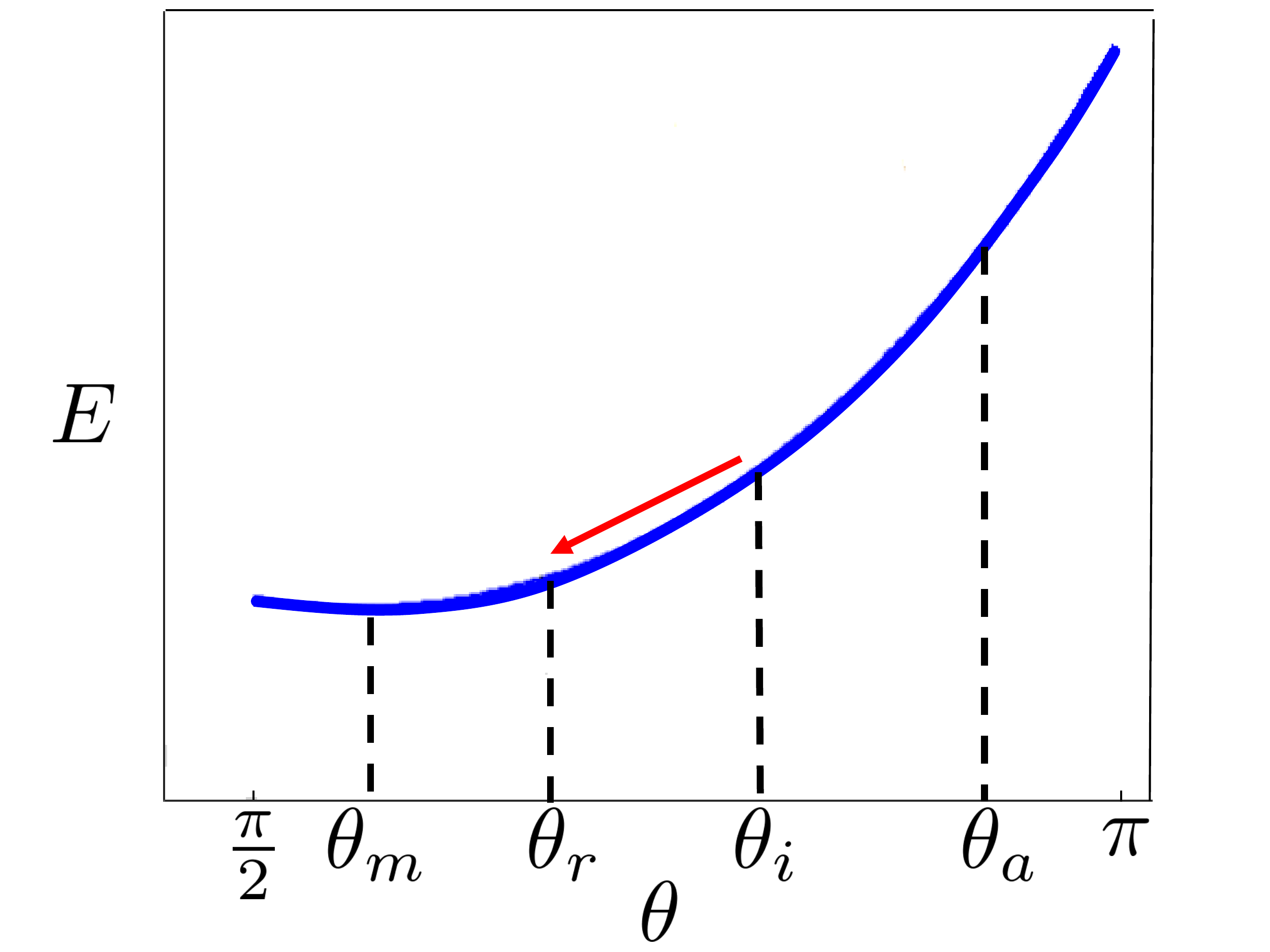}
\caption{Case $\theta_{m} < \theta_{r}$. The contact angle  decreases, while the opening angle increases, and $\theta$ reaches $\theta_{r}$, after which the contact line moves until $\alpha = -\pi/2 + \theta_{r}$ is reached.
The initial folding angle was $\alpha_{i}={\pi}/{18}$ (10$^\circ$), the initial contact angle $\theta_{i}={3\pi}/{4}$ (135$^\circ$), and $\theta_{E}={3\pi}/{4}$ (135$^\circ$). {Note that the  chosen values for $\theta_a$ and $\theta_r$  are arbitrary and the shape of the energy profile shown here is generic}.}
\label{fig:hyst1}
\end{figure}

We first consider the case for which $\theta_{m} < \theta_{r}$. Physically, this is the situation in which one starts with a folding angle which is too small and the system needs to open up (increase in $\alpha$) to reach the minimum of surface energy. 
With the initial conditions and the information on the ratio of the surface energies provided by $\theta_{E}$, we can calculate the profile of energy for a fixed contact line, with  results  shown in  Fig.~\ref{fig:hyst1}. Starting from the initial position,  $\theta$ first decreases until it reaches $\theta_{r}$, which is accompanied by an increase of the  folding angle. Once  the receding angle is reached, the contact line starts moving. The energy minimization problem changes and becomes an evolution at fixed contact angle $\theta = \theta_{r}$. Neglecting the dynamics and assuming a quasi-static evolution, this is similar to  the problems solved in \S\ref{2D}. The motion of the contact line stops as soon as the equilibrium  folding angle is reached, and we get therefore folding with  $\alpha = -\pi/2 + \theta_{r}$ by force balance (at that point we therefore have $\theta_m=\theta_r$).

\subsection{$\theta_{m} > \theta_{a}$}

The other relevant limit to consider is the one in which $\theta_{m} > \theta_{a}$ which corresponds to a situation in which one starts with a folding angle which is too large. This is the situation relevant to experiments where, typically, the initial fabrication stage  initially leads to planar structures with $2\alpha=\pi$ which will be folded by surface tension into  three-dimensional shapes \cite{leong07,leong08}. In this case, the dynamics is similar to the previous case with the receding angle replaced by the  advancing angle. With the contact line pinned, the contact angle increases (and the folding angle decreases) until $\theta_a$ is reached. At this point, the contact moves with 
$\theta = \theta_{a}$ fixed, until the system reaches its local minimum of surface energy. The final folded configuration is thus  characterized by the angle $\alpha = -\pi/2+\theta_{a}$.

\subsection{Configuration hysteresis}

\begin{figure}[t]
\includegraphics[width=.43\textwidth]{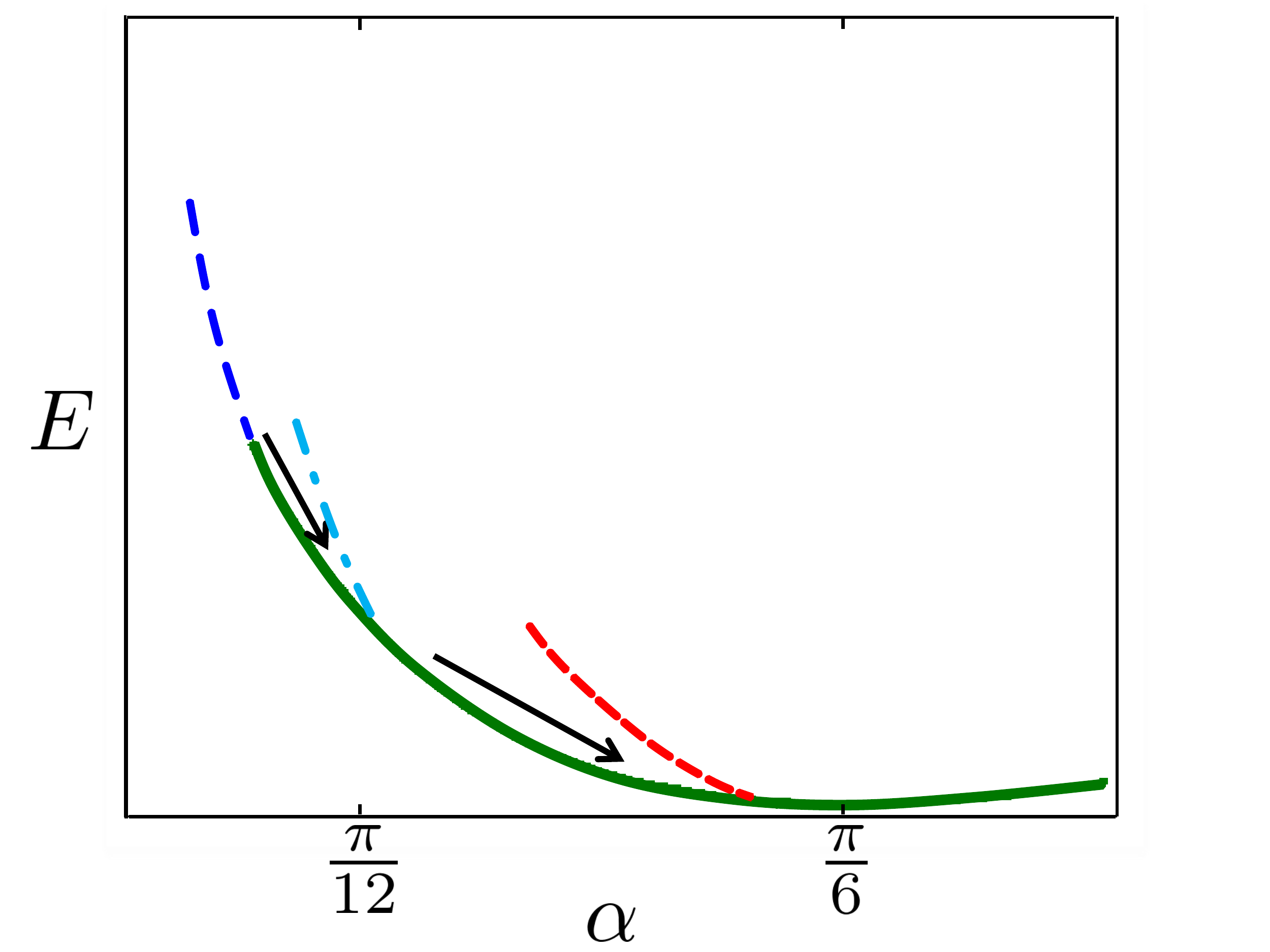}
\quad
\includegraphics[width=.45\textwidth]{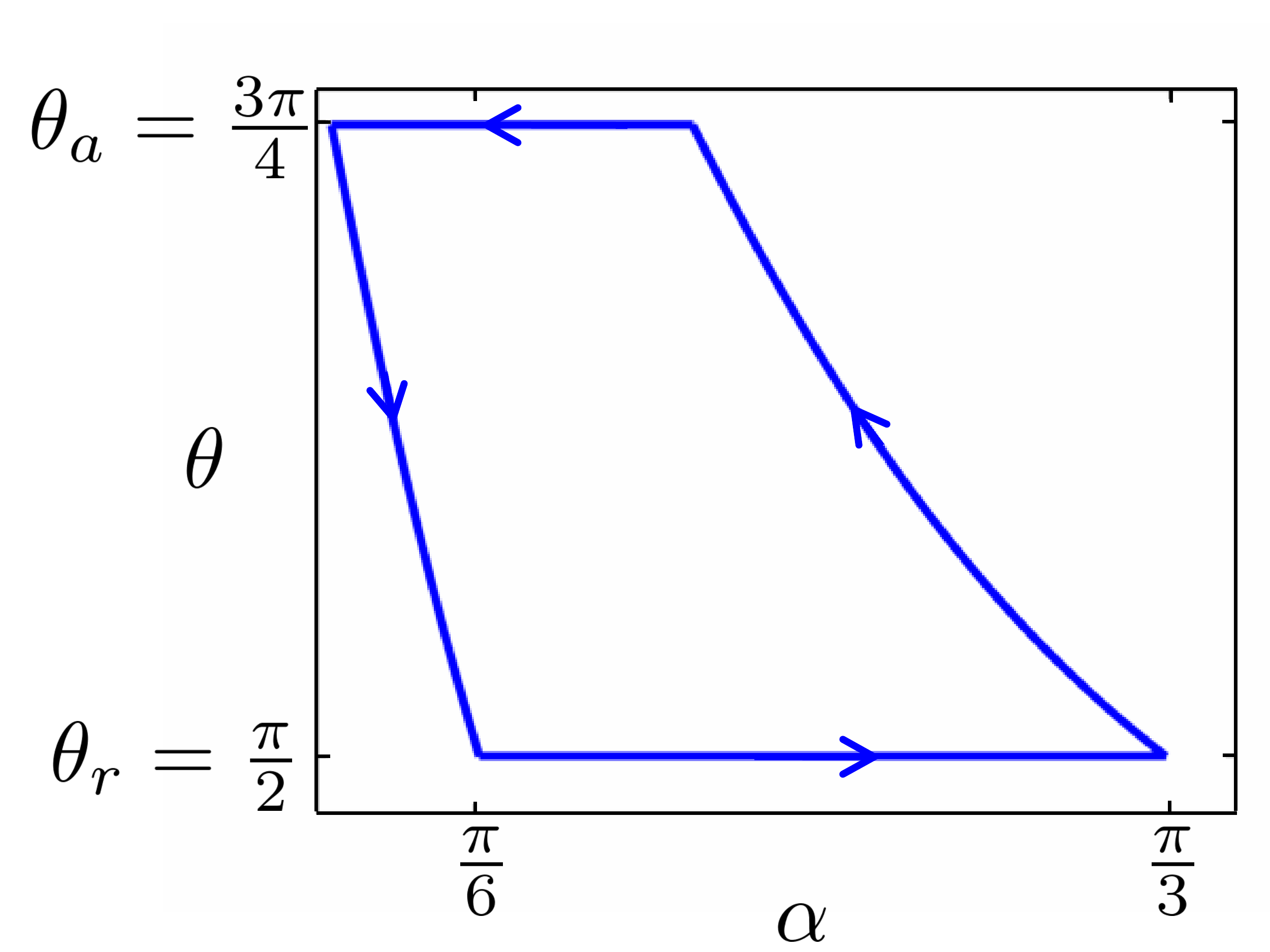}
\caption{Left: Evolution from a folded configuration (small value of $\alpha$). Surface energy, $E$, as a function of the  folding angle $\alpha$. The value of  $\alpha$ increases with $\theta=\theta_{r}$ fixed until the configuration with a local minimum of energy is reached (solid line). If the system goes backward and the hinge closes, $\theta$ is no longer constant but the contact line is pinned, and the surface energy goes back up but along a different path (dashed lines). Right: Hysteresis for the contact angle as a function of the folding angle  if we open and close successively the system ($\theta_{r} = {\pi}/{2}$, $\theta_{a} = {3\pi}/{4}$).}
\label{fig:hyst2}
\label{fig:summary_hysteresis}
\end{figure}

Contact angle hysteresis leads to non-reversible paths in the configuration space. To illustrate this, let  us imagine that we can  increase or decrease manually the folding angle at the point where $\theta=\theta_{r}$. If we try to decrease it, we will come back to an evolution at fixed contact line with an increase in  $\theta$. This is shown in Fig.~\ref{fig:hyst2} (left) with $\theta_{r}={2\pi}/{3}$. If at a given point we go backward and  decrease the folding angle, the energy profile encountered by the system is the one in dashed lines, representing the energy for an evolution at fixed contact line, which is different from the solid line showing an evolution at fixed (receding) contact angle. This can be further illustrated by imagining a controlled experiment where the system is successively opened and closed (so $\alpha$ moves backward and forward between two values), with the hysteresis leading to net work being performed on the system.   
If we then sketch the value of the contact angle, $\theta$, as a  function of the folding angle,  $\alpha$, we get a  typical hysteretic  path (Fig.~\ref{fig:summary_hysteresis}, right). 

\subsection{De-wetting and contact angle hysteresis}

In \S\ref{2D} we addressed the issue of de-wetting near the hinge and showed that a range of  folding angles was leading to the same value of the surface energy.  How is this changed by contact angle hysteresis?   It is straightforward to see that  contact angle hysteresis will actually prevent de-wetting altogether. The simplest argument is energetic. We saw in \S\ref{2D} that, in the case of a single contact angle, both wetted and de-wetted configurations had the same surface energy. With hysteresis this is no longer the case. In order to de-wet near the hinge point, energy must be supplied to the system in order to induce a motion of the contact line. Starting with one of the three final configurations outlined in \S\ref{hyst}A-C which correspond to local minima of surface energy, external work would be required in order to  move the new free surface away from the hinge point   and, if the external contact angle is $\theta=\theta_a$ (situation outline in \S\ref{hyst}C) move the external contact line as well. De-wetting will therefore not occur spontaneously in this system.

\section{Folding in three dimensions}\label{3D}
\subsection{Idealized geometry: infinite folding}
\label{3Dinfinite}

We now consider the three-dimensional case, and assume for simplicity that we have no contact angle hysteresis. The main complication going from two to three dimensions is the fact that the shape is no longer a simple circular arc, but a more complex  surface of constant mean curvature.

\subsubsection{Hydrophobic case}

\begin{figure}[t]
\includegraphics[width=.45\textwidth]{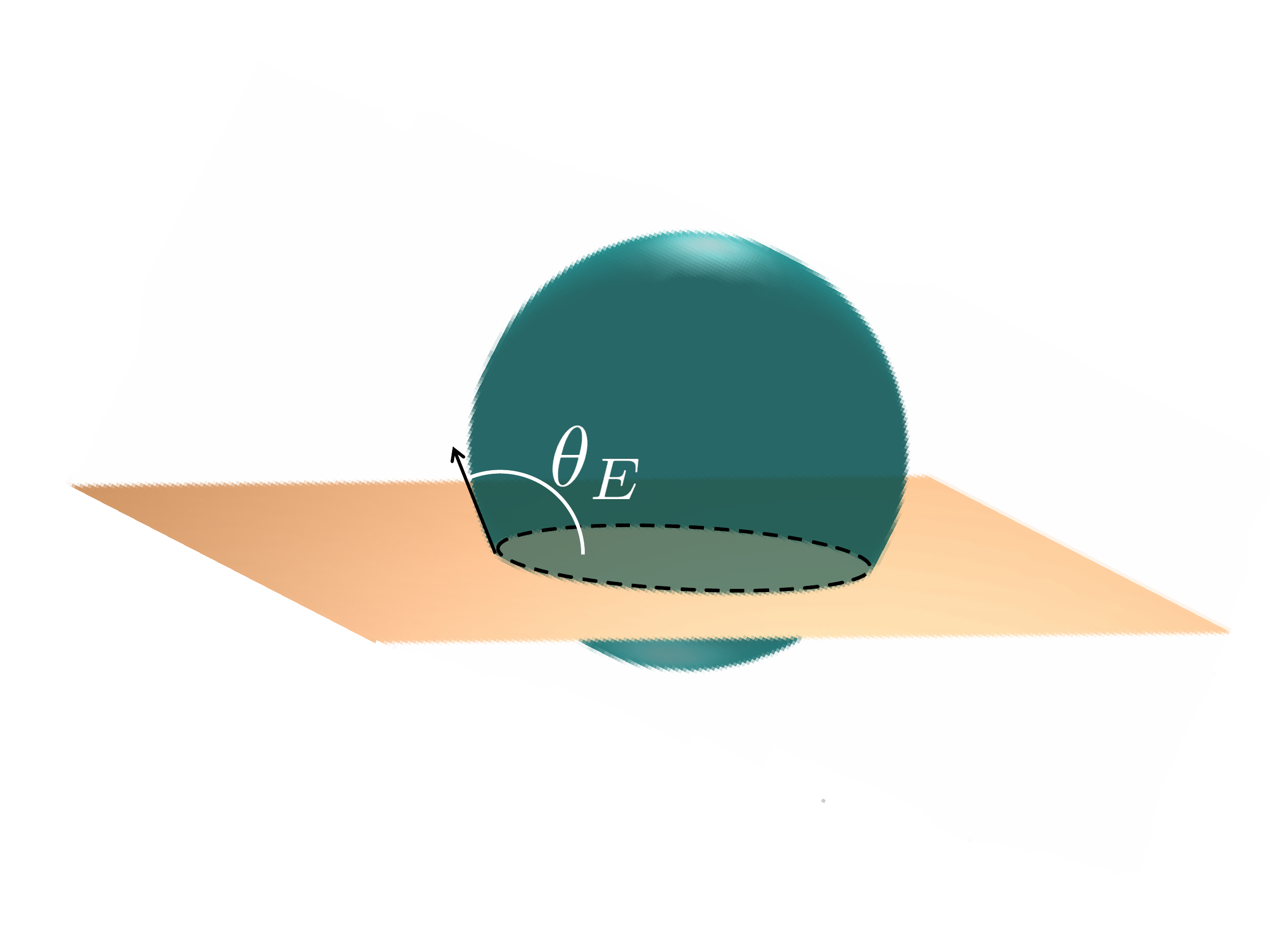}
\quad\quad
\includegraphics[width=.36\textwidth]{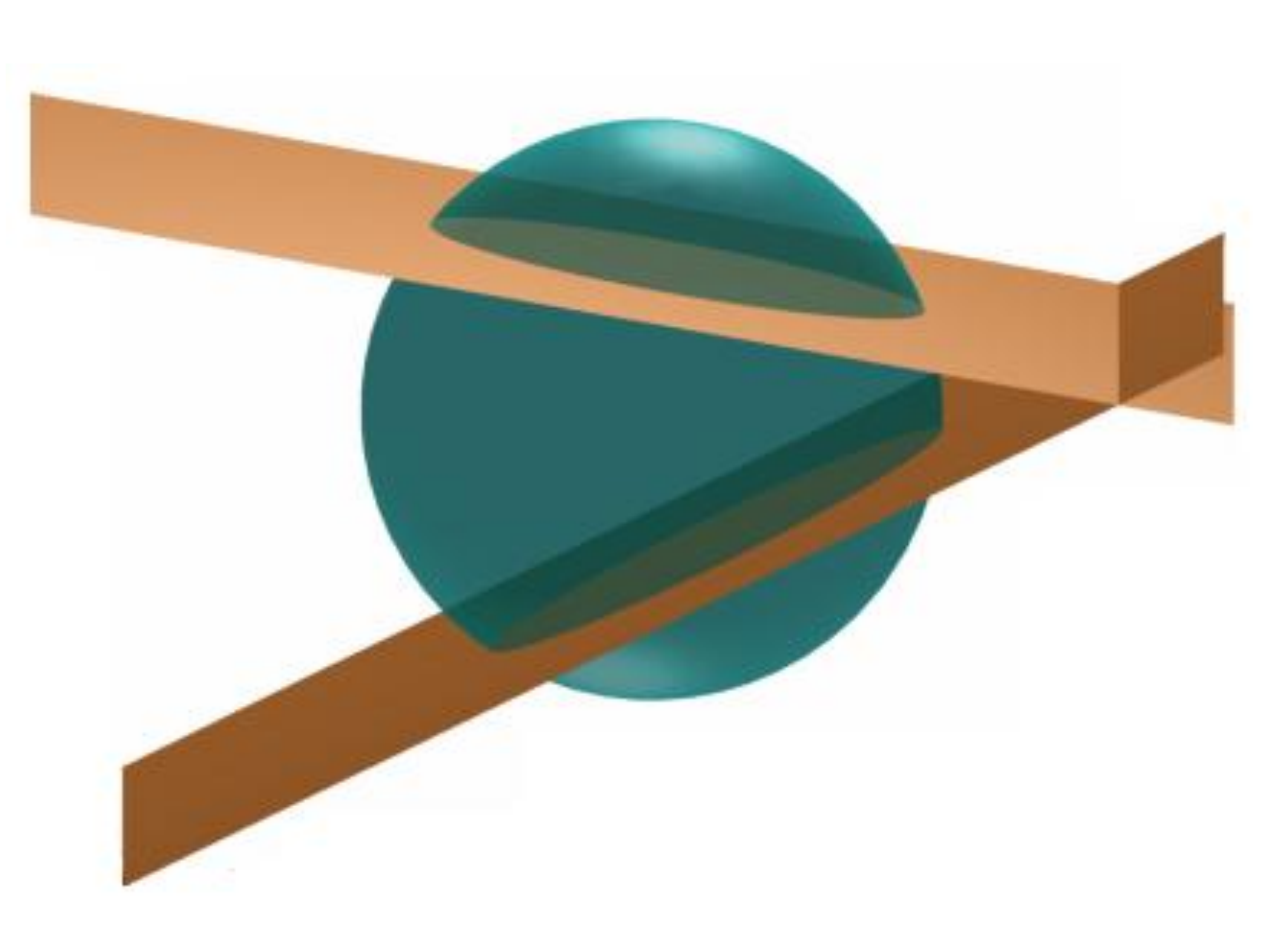}
\caption{Left: The intersection between a sphere and a planar surface leads to a constant contact angle along the intersection line. Right: Two planar surfaces intersecting a sphere and forming a corner satisfy the contact angle condition along both contact lines. Removing  the caps  leads to  the actual droplet configuration.}
\label{fig:plan_sphere}
\label{fig:2plans_sphere}
\end{figure}

\begin{figure}[t]
\includegraphics[width=.5\textwidth]{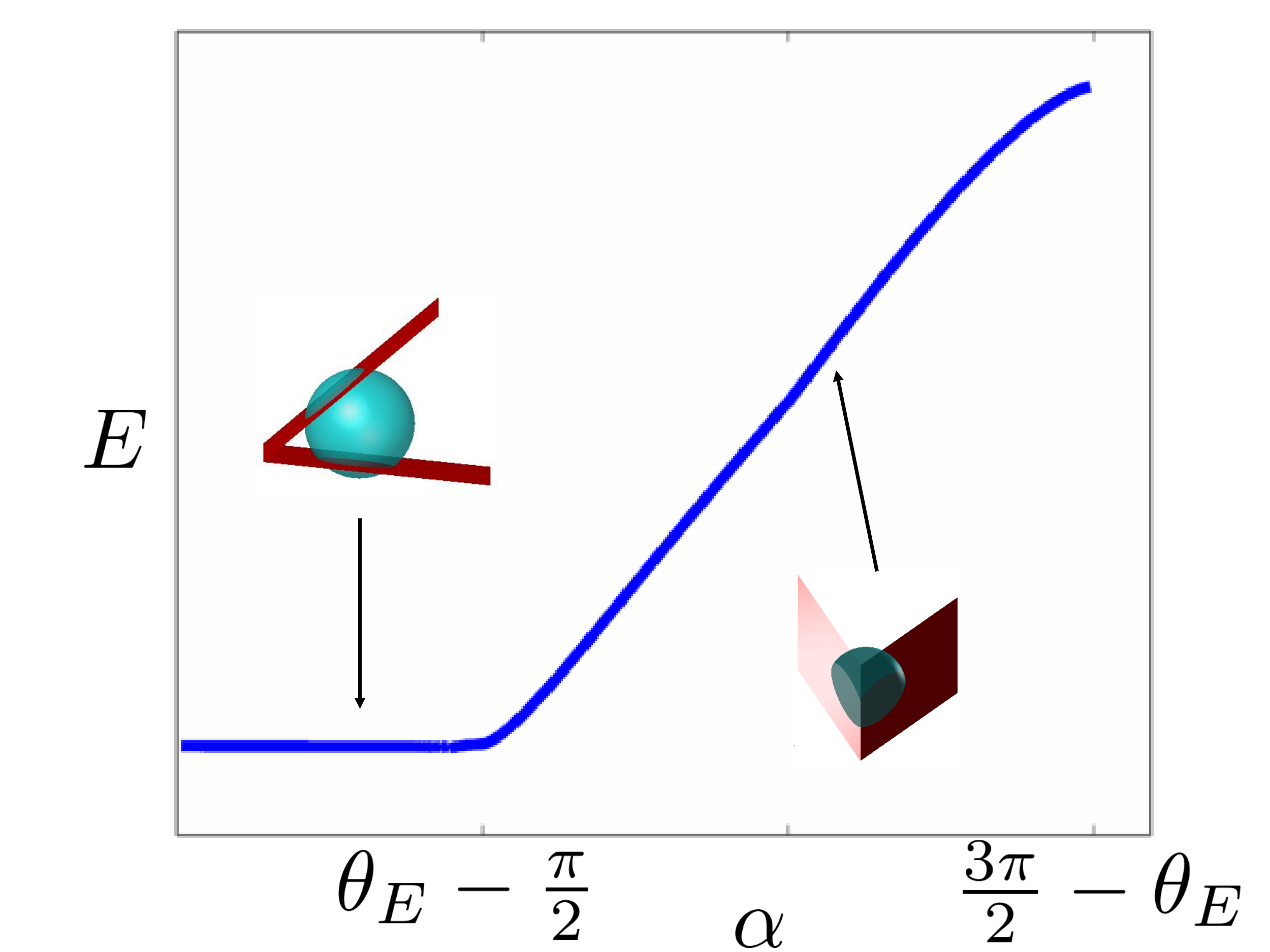}
\caption{Total surface energy, $E$, as a function of the folding angle, $\alpha$, in three dimensions, for  $\theta_E={3\pi}/{4}$. For $\alpha  \leq  -{\pi}/{2} + \theta_E$, the energy does not depend on $\alpha$,  as in   two dimensions.}
\label{fig:plot_dewet_3D}
\end{figure}

The droplet shape must satisfy three constraints: (1) constant volume, (2) constant contact angle, and (3) constant mean curvature. Clearly, the portion of a sphere is a shape of constant mean curvature. Let us show that this shape can also satisfy the other two constraints. Let us imagine that we make a planar surface intersect a sphere. The contact angle between the sphere and the plan will be constant over the intersection line (see Fig.~\ref{fig:plan_sphere}, left). By adding another surface intersecting the sphere with the same angle, and by removing both  spherical caps  on the other side of the surfaces, we get a configuration matching both the constant contact angle and the constant mean curvature conditions (Fig.~\ref{fig:2plans_sphere}, right). Changing the radius of curvature of the droplet allows to tune the volume, and thus the spherical solution allows to satisfy all three constraints. Note that this solution is unique since the spherical shape ensures that the  energy of the free surface is minimized. It is worth mentioning that if the folding angle $\alpha$ is between 0 and $-\pi/2 +\theta_E$, the intersection line between the two planes does not intersect the sphere, corresponding thus to the de-wetting configuration.  With the solution known for the droplet shape, we can compute the surface areas giving the surface energies as a function of the folding angle, with results shown in  Fig.~\ref{fig:plot_dewet_3D}. Similarly to the two-dimensional case and for similar geometrical reasons, the surface energy stays constant in the de-wetting range ($\alpha $ between 0 and  $-\pi/2 +\theta_E$).

\subsubsection{Hydrophilic case}

\begin{figure}[t]
\includegraphics[width=.5\textwidth]{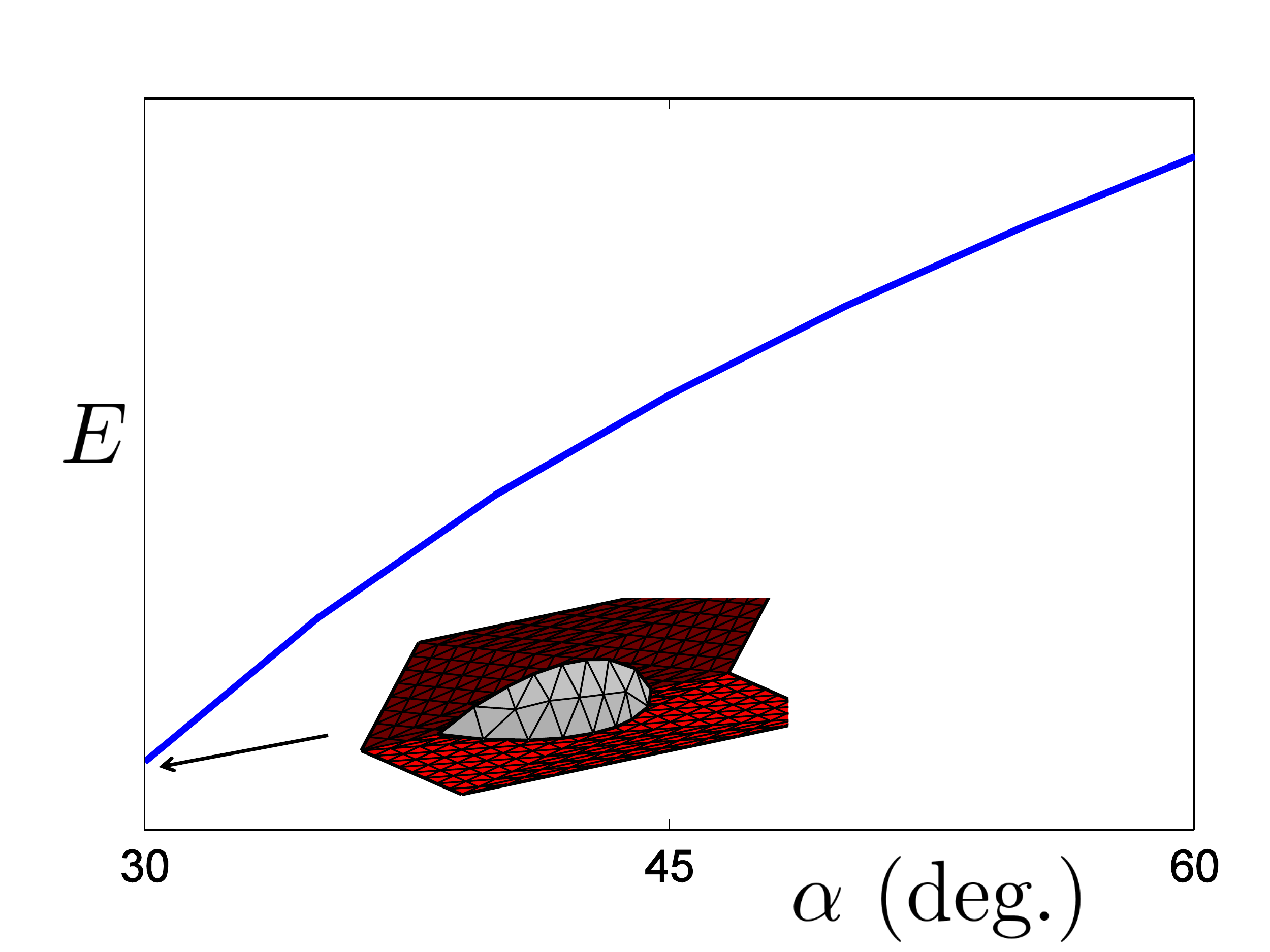}
\caption{In three dimensions, and similarly to  two dimensions, the energy increases with $\alpha$ in the hydrophilic case (computations with Surface Evolver). The system is driven towards $\alpha = 0$}
\label{fig:SEhydrophilic}
\end{figure}

In the hydrophilic case, computing the  shape of the droplet is more difficult. Given that two and three dimensions showed similar results in the hydrophobic case, we can conjecture  that the same occurs in the hydrophilic limit and that  complete folding will be the final configuration. This hypothesis is confirmed by numerical simulations  performed using Ken Brakke's Surface Evolver \cite{brakke92} and shown in Fig.~\ref{fig:SEhydrophilic}. The total surface energy monotonically decreases when the folding angle decreases, and complete folding leads to the global minimum. 

\begin{figure}[b]
\includegraphics[width=.4\textwidth]{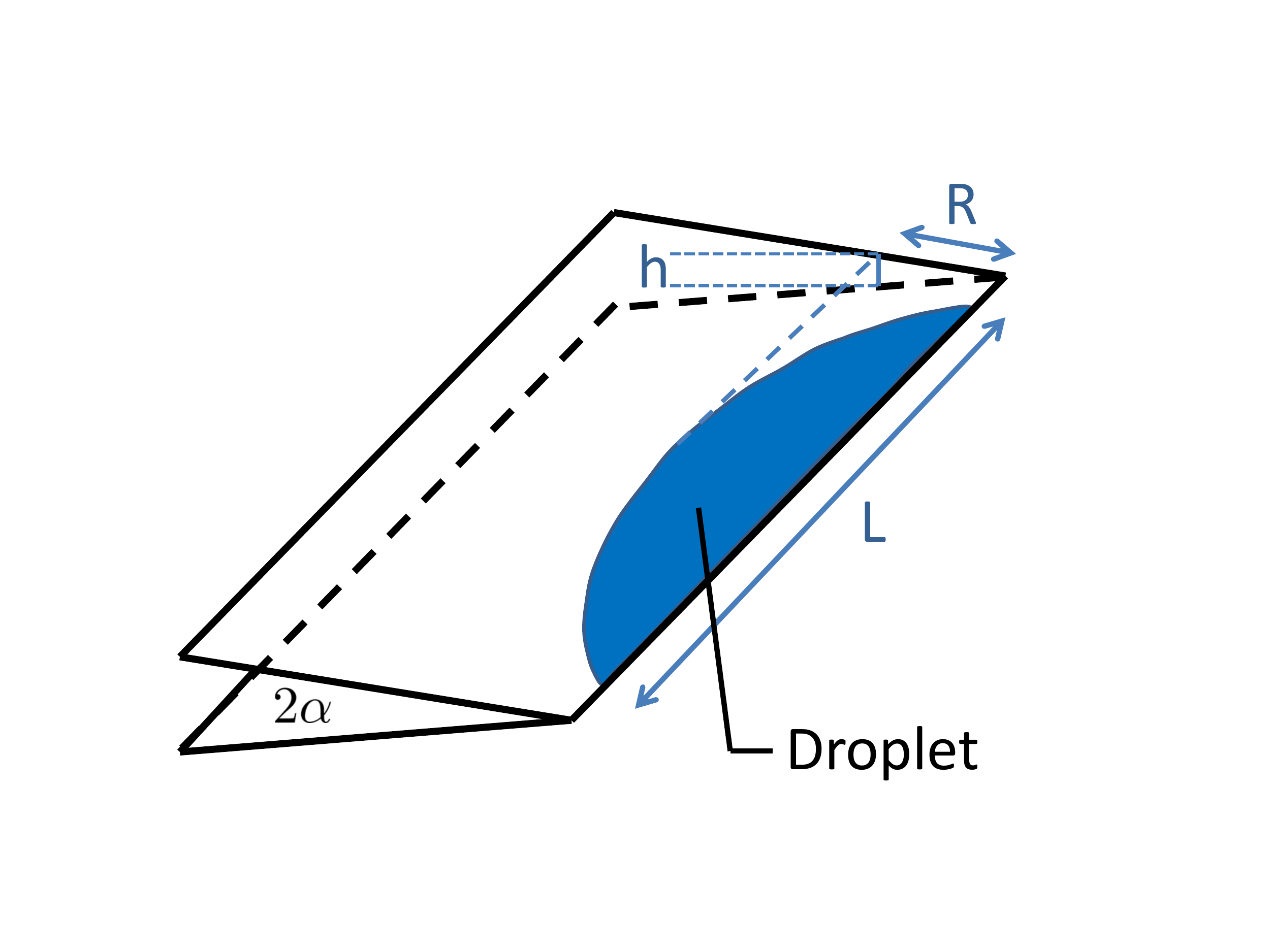}
\caption{In the hydrophilic case, for small values of (and changes in) $\alpha$ we assume that the droplet remains of a similar shape (see text for notation).}
\label{fig:hydrophilic-3D}
\end{figure}

This situation can be further  addressed using scaling analysis, similarly to what was done in two dimensions.   Let us assume that, for small variations of the folding angle $\alpha$, the droplet  shape is not modified and can be parametrized as in Fig.~\ref{fig:hydrophilic-3D}, with $L = BR$, $B$ being a constant shape parameter. For small values of $\alpha$, the opening $h$ can be approximated as $h\sim R\alpha$. The volume of the droplet, $V$, is then proportional to $V\sim RLh = BR^{3}\alpha$, and the surface energy can be approximated by 
\begin{equation}
E \sim 2C_{1}RL(\gamma_{SL}-\gamma_{AS}) + C_{2}Lh\gamma_{AL},
\end{equation}
 where $C_{1}$ and $C_{2}$ are two other shape constants related to the dimensionless areas of the solid-liquid and  liquid-air interfaces respectively. Replacing $R$ by its expression given by $V$ and $\alpha$ yields 
\begin{equation}\label{above}
E \sim 2BC_{1}(\gamma_{SL}-\gamma_{AS}) \left( \frac{V}{B \alpha} \right)^{\frac{2}{3}}+B\gamma_{AL} C_{2} \alpha^{\frac{1}{3}} \left(\frac{V}{B} \right) ^{\frac{2}{3}}.
\end{equation}
Differentiating $E$ from Eq.~\eqref{above} and using Young's formula to relate the surface energies leads to
\begin{equation}
\frac{{\rm d}E}{{\rm d}\alpha} \sim B\gamma_{AL}\left( \frac{V}{B} \right)^{\frac{2}{3}} \left(\frac{4}{3}C_{1} \cos(\theta_{E}) \alpha^{-1}+\frac{1}{3}\right) \alpha^{-\frac{2}{3}}.
\end{equation}
Since in the hydrophilic case $\cos\theta_{E} >0$, we have ${{\rm d}E}/{{\rm d}\alpha}>0$ for all values of $\alpha$, and therefore the minimum of energy is obtained for $\alpha=0$. As in the two-dimensional case, a hydrophilic droplet in three dimensions leads to  complete folding.

\subsection{Finite-size folding in three dimensions}
\label{3Dfinite}

\begin{figure}[t]
\includegraphics[width=.45\textwidth]{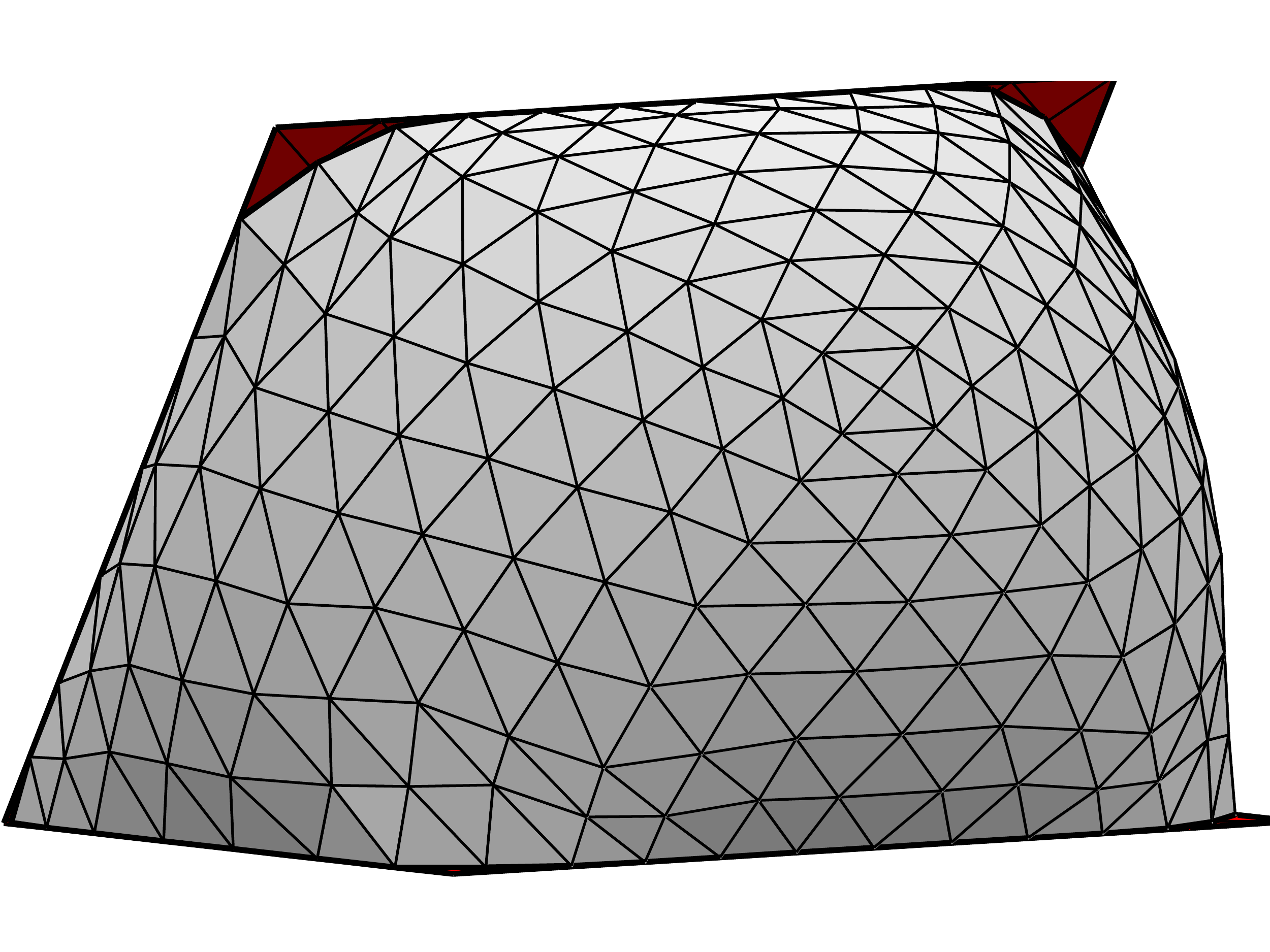}
\caption{Example of final three-dimensional folded  configuration visualized with Surface Evolver, assuming a contact angle $\theta_E=\pi/2$.  The size of the square supports is $1\times 1$ and the volume of the droplet 0.5. The edges deform the otherwise spherical shape.}
\label{fig:SE_90degrees}
\end{figure}

As we did in \S\ref{2D}B, we now  examine the three-dimensional folding  behavior as the interface reaches the edges of the wall. In that case, as a difference with the easier  two-dimensional case, the droplet edges deform the spherical cap shape, and an analytical treatment is difficult. We use Surface Evolver \cite{brakke92} to analyze  the system behavior, illustrating the results with $\theta_{E}=\pi/2$, and $\theta_{E}=3\pi/4$. The goal is to investigate if the results are qualitatively similar to those obtained in two-dimensions (as we will see below, they are). 

\begin{figure}[t]
\includegraphics[width=.65\textwidth]{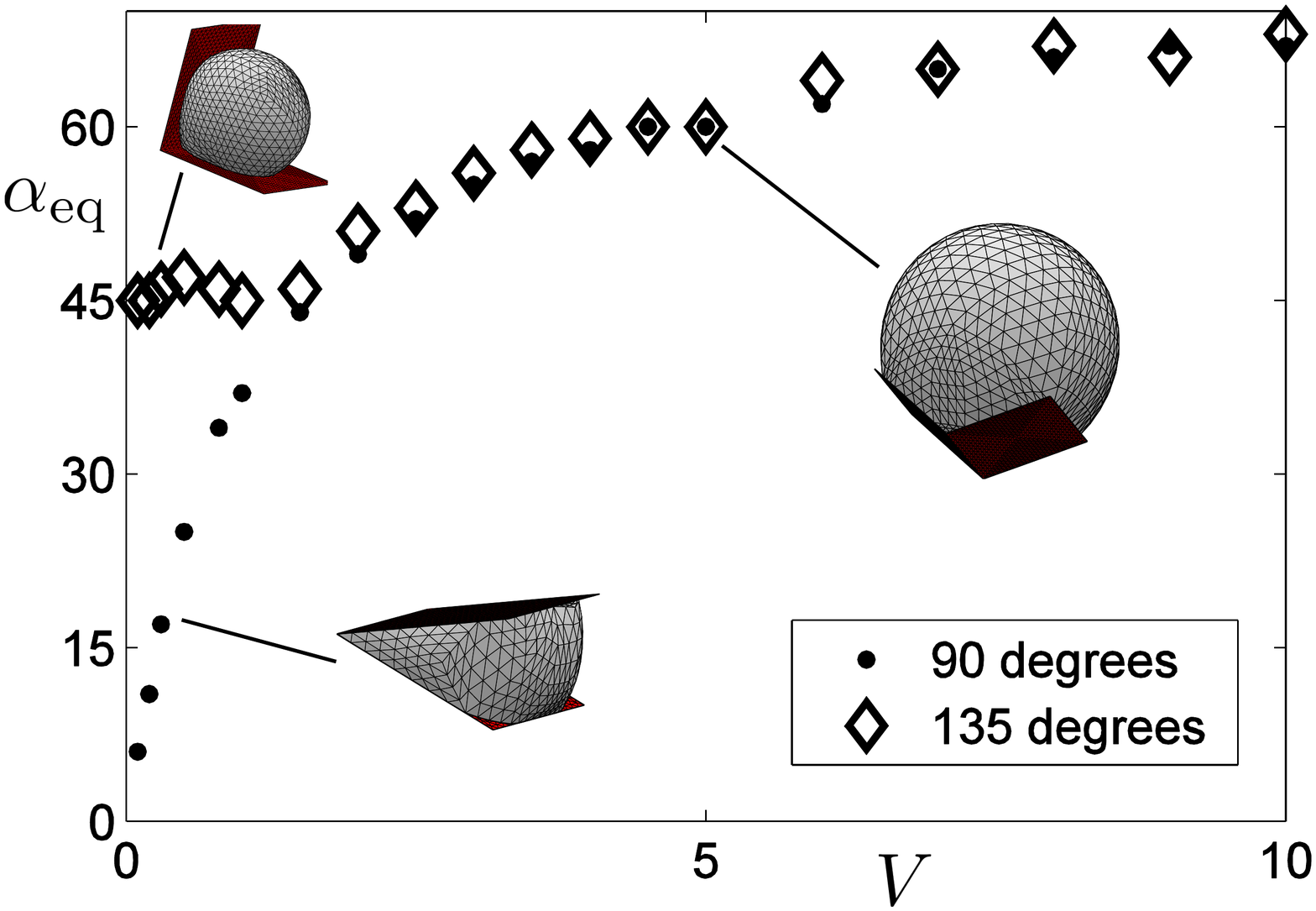}
\caption{Folding angle, $\alpha_{\text{eq}}$,  vs.~droplet volume, $V$, for finite-size three-dimensional folding,  obtained numerically using Surface Evolver. The behavior is  similar to the two-dimensional calculations (Fig.~\ref{fig:pb_au_bord}). For $\theta_E=3\pi/4$, $\alpha_\text{eq}=\pi/4$ and is independent of volume until the droplet reaches the edges of the wall. After that stage, the folding angle increases with the droplet volume. For $\theta_E=\pi/2$, the edges are reached for all volumes, and the folding volume always increases with the droplet volume. In both cases, $\alpha_\text{eq}$ slowly asymptotes to  $\pi/2$ as $V\to\infty$.}
\label{fig:plot_evolver}
\end{figure}

Simulations are performed with two square walls of size $1\times 1$ attached by a hinge. We fix the contact angle, vary the folding angle, and for each value of $\alpha$ compute the total surface energy, allowing us to determine the particular value $\alpha_\text{eq}$ at which the surface energy is minimum. Final numerical results are then converged to through mesh refinements. As an example, we illustrate in Fig.~\ref{fig:SE_90degrees} the shape of the droplet at equilibrium  in the case of $\theta_E=\pi/2$ and $V=0.5$ (leading to $\alpha_\text{eq}=25^\circ$).

We display the results for the folding angle as a function of the droplet volume  in Fig.~\ref{fig:plot_evolver}. They are close to what was obtained in two dimensions (Fig.~\ref{fig:pb_au_bord}). In the case where $\theta_E=3\pi/4$, $\alpha_\text{eq}$ remains constant and equal to  $\pi/4$ ($=-\pi/2 + \theta_E$) until the droplet  reaches the edges of the wall.
As  in two dimensions, the folding angle remains therefore independent of the droplet volume until that point. After reaching the critical volume, the equilibrium angle increases with the volume. In the case $\theta_E=\pi/2$, the edges are reached for any small droplet volume since  the infinite-wall case would lead to a completely folded configuration,  $\alpha_\text{eq}=-\pi/2 + \theta_E=0$. The  folding angle increases therefore monotonically with the droplet volume in that case.  Note that for both contact angles, $\alpha$  reaches $\pi/2$   asymptotically  in the limit of  large volumes. 

\section{Folding of a two-dimensional elastic sheet}\label{elastic}

Instead of considering two rigid walls joined at a free hinge, we add in this section a new ingredient, namely the elastic cost of deforming the solid. We adopt the same setup as the one studied in Ref.~\cite{py07} in the context of capillary origami, namely a purely two-dimensional configuration of an elastic sheet bent by a two dimensional droplet. In the study of Ref.~\cite{py07}, the droplet was assumed to always reach the edge of the sheet. As a difference, we will focus here on the case in which the droplet does not reach the edge of the sheet, and investigate the role of wetting on the equilibrium folding geometry. 

\subsection{Geometry and parameters}

\begin{figure}[t]
\includegraphics[width=.44\textwidth]{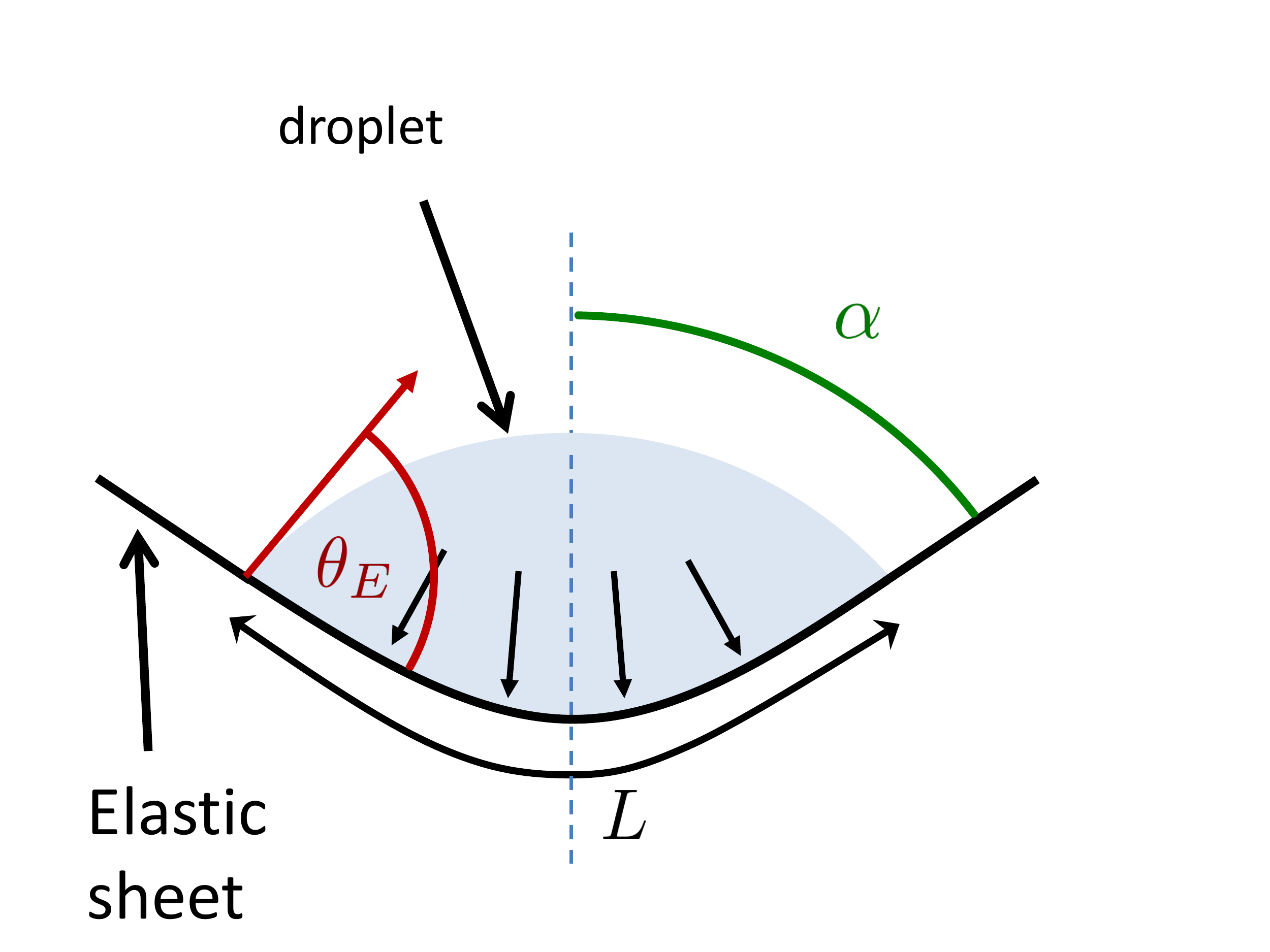}
\quad\quad\quad
\includegraphics[width=.42\textwidth]{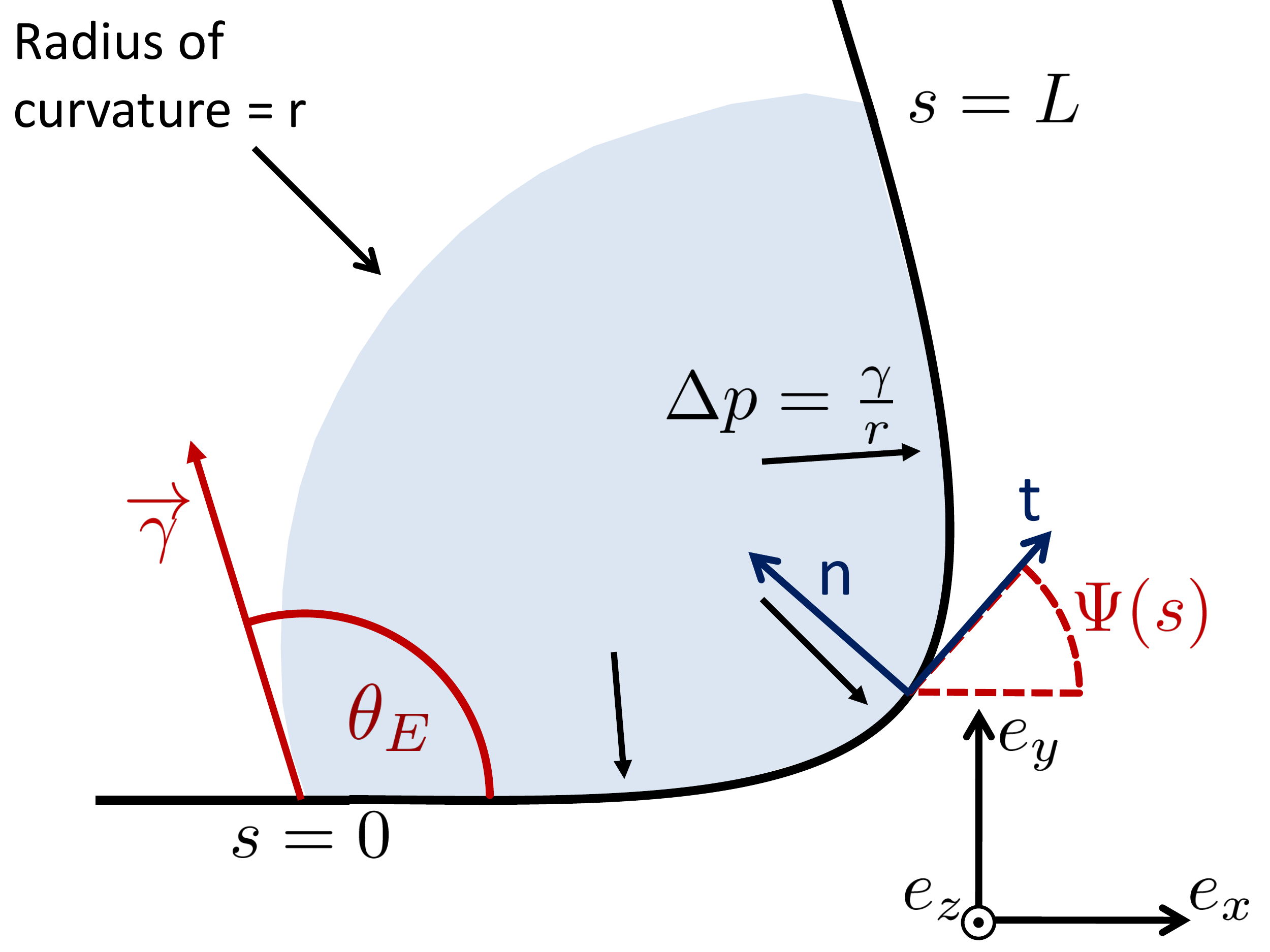}
\caption{Elastic sheet bent by a droplet in two dimensions. Left: The contact angle is $\theta_{E}$ and the folding angle $\alpha$. The droplet wets a  total length $L$ along the rod. Right: specific notation used to enforce force and moment balance along the flexible sheet (see text for details). 
}
\label{fig:flexible}
\end{figure}

The problem geometry and its parameters are illustrated in Fig.~\ref{fig:flexible}. As in the previous sections, the contact angle is denoted $\theta_E$ and we ignore the effects of gravity. The droplet is assumed to wet a (two-dimensional) length $L$. The portion of the sheet not wet by the droplet remains straight (since they are subject to no external force or moment) and can be used to define the folding angle $\alpha$. By analogy to the previous problems, $\alpha$ is half the angle formed by these lines. A flat sheet with no bending corresponds thus to a  folding angle of  $\pi/2$.  The goal of this section is to compute the folding angle as a function of the (two-dimensional) volume of the droplet and its contact angle.

The notation for the model are detailed in Fig.~\ref{fig:flexible}, right. 
We denote by \textbf{F} the resultant internal stress acting on a cross-section of the sheet, \textbf{K} the external force per unit length, \textbf{M} the resultant internal moment, and \textbf{t} the unit tangent vector to the rod representing the sheet. We parametrize the rod with the curvilinear  coordinate $s$, with $s=0$ set at one of the two triple points and $s=L$ at the other. With this variable,  the complete set of equilibrium equations for the sheet are written as  \cite{landau_elasticity}
\begin{equation}
\frac{d\textbf{F}}{ds}=-\textbf{K}, \quad  \frac{d\textbf{M}}{ds}=\textbf{F} \times \textbf{t}.
\label{eq:rod}
 \end{equation}

Let us call $\psi$ the angle between the tangent of the sheet (represented by \textbf{t}) and the $x$ axis. We know that, in this problem, $\textbf{K}=-\Delta p\,\textbf{n}=-({\gamma}/{r})\,\textbf{n}$, where $r$ is the radius of curvature of the droplet and  $\gamma$ the surface tension.  Assuming linear elasticity, the equilibrium condition, Eq.~\eqref{eq:rod}, must now be closed  by the Hookean constitutive relationship relating the magnitude of the bending  moment,  $M$, to the sheet curvature as \cite{landau_elasticity}
 \begin{equation}\label{Hook}
M=EI\frac{d\psi}{ds}, 
\end{equation}
where $EI$ is the sheet bending stiffness.  Written in the Frenet-Serret frame $\{ {\bf t},{\bf n}\}$, mechanical equilibrium  on the interval  $0 < s < L$  takes the form 
\begin{subeqnarray}
 \frac{dF_{n}}{ds}+F_{t}\frac{d\psi}{ds}&=&\frac{\gamma}{r},  \\
 \frac{dF_{t}}{ds}-F_{n}\frac{d\psi}{ds}&=&0, \\
 \frac{dM}{ds}&=&-F_{n},
 \label{eq:rod_expanded}
 \end{subeqnarray}
together with Eq.~\eqref{Hook}. 

The system in Eq.~\eqref{eq:rod_expanded} needs to be accompanied by appropriate force boundary conditions,  which can simply be found by projecting the surface tension in the directions tangent and normal to the sheet at the triple point, 
\begin{subeqnarray}
  F_{n}(0^{+}) &=& -\gamma \sin \theta_E,  \\
  F_{n}(L^{-}) &=& \gamma \sin \theta_E, \\
  F_{t}(0^{+}) &=& -\gamma \cos \theta_E, \\
  F_{t}(L^{-}) &=& -\gamma \cos \theta_E. 
 \label{eq:boundary_conditions}
 \end{subeqnarray}

The value of $L$ is, however, unknown and must be computed. It can be calculated  by solving the equation with the boundary condition at $s = 0$ and finding the location along the sheet where the values of the stresses match the force boundary conditions. It is to be noted also that with this set of equations we cannot enter the volume directly as an input. Instead, we  set a value for the radius of the droplet, $r$,  solve the equations, and  get both the volume and the folding angle {\it a posteriori}. If we span all the possible values for $r$, then we can obtain the folding angle as a function of the droplet volume. 

\subsection{Solving the problem}

We start by re-arranging Eq.~(\ref{eq:rod_expanded}) by differentiating once to eliminate $F_t$, $F_n$, and $M$  and obtain a third-order equation for $\psi$
  \begin{equation}
\frac{d^{3}\psi}{ds^{3}}+\frac{\gamma \cos \theta_E}{EI}  \frac{d\psi}{ds}+\frac{1}{2}\left(\frac{d\psi}{ds}\right)^{3}+\frac{\gamma}{EIr}=0
\label{eq:dimensional_eq},
\end{equation}
valid for $0 < s < L$. 
We nondimensionalize Eq.~(\ref{eq:dimensional_eq}) by introducing the {elasto-capillary length} $L_{EC}=\left({EI}/{\gamma}\right)^\frac{1}{2}$ introduced in Refs.~\cite{py07,bico04,py07_EPL}. If the radius of curvature of the droplet is $r$ and the curvilinear coordinate is $s$,  we define the dimensionless radius $\overline{r}={r}/{L_{EC}}$ and the reduced curvilinear abscissa $\overline{s}={s}/{L_{EC}}$. The dimensionless third order differential equation is then
 \begin{equation}
\quad \frac{d^{3}\psi}{d\overline{s}^{3}}+\cos \theta_E  \frac{d\psi}{d\overline{s}}+\frac{1}{2}\left(\frac{d\psi}{d\overline{s}}\right)^{3}+\frac{1}{\overline{r}}=0,
\label{eq:adim}
\end{equation}
for $0 < s < L/L_{EC}$. Using ' to denote derivatives with respect to $s$, the boundary conditions are  $\psi(0)=0$ (arbitrary condition to set the origin of the coordinate system), 
$\psi ' (s=0)=0$ (no moment condition)  and 
$ \psi'' (s=0)=\sin \theta_E $ (force condition).
\begin{figure}[t]
\includegraphics[width=.5\textwidth]{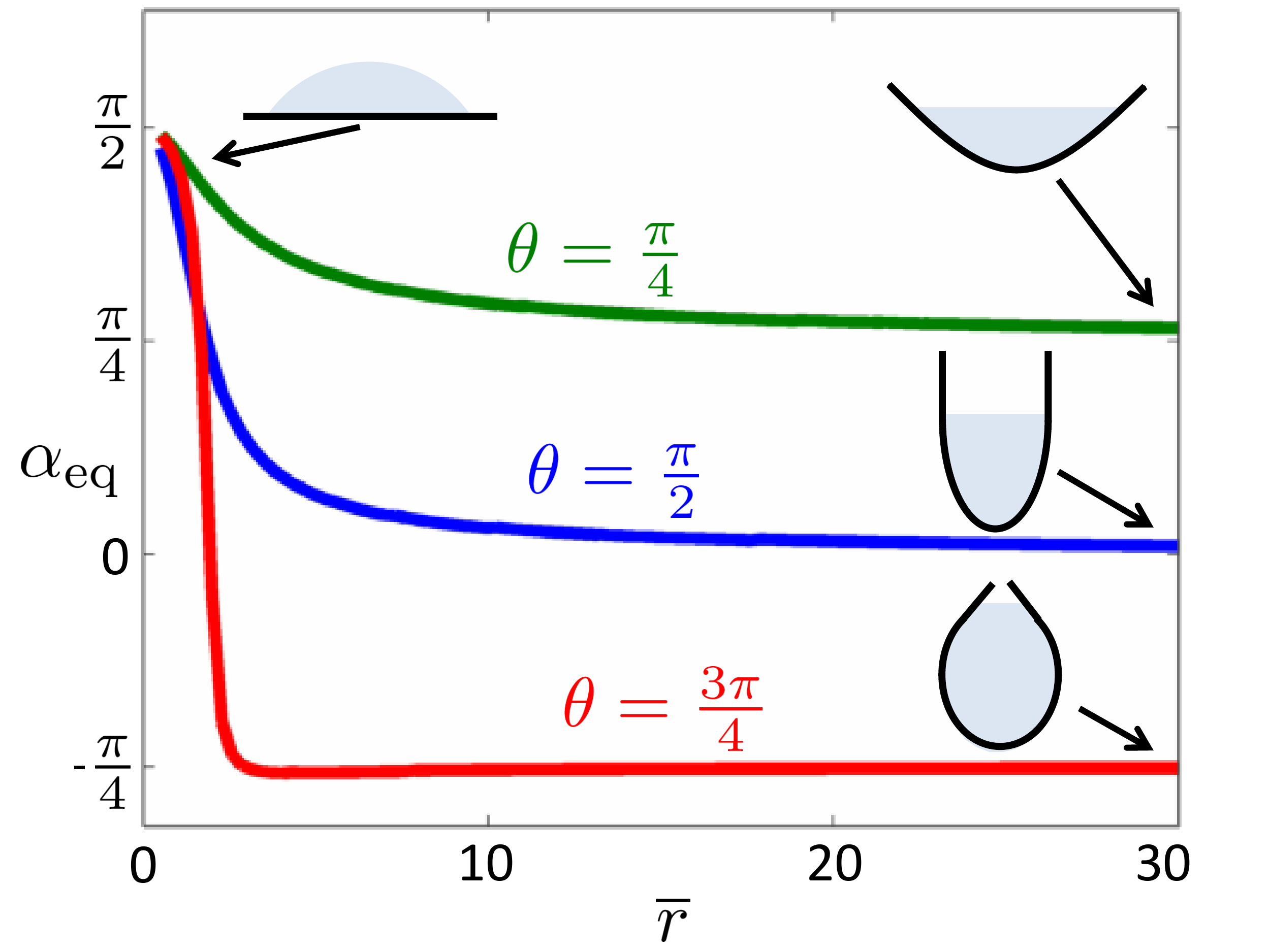}
\caption{Folding angle, $\alpha$, as a function of the non-dimensional droplet curvature, $\overline{r}$. For small $\overline{r}$ (large stiffness), $\alpha$ is close to ${\pi}/{2}$, and there is no  folding. In the opposite small-stiffness limit of large  $\overline{r}$, $\alpha$ tends to ${\pi}/{2}-\theta_E$.}
\label{fig:alpha_against_r}
\end{figure}

We  compute the solution to Eq.~\eqref{eq:adim} numerically using Matlab. We find the missing dimensionless parameter $L/L_{EC}$ by noticing that it corresponds to the  positive value of the curvilinear coordinate $s$ for which $\psi'(\overline{s}={L}/{L_{EC}})=0$ (no moment at the exit contact line). Once  $\psi$ and $L/L_{EC}$ are known, we can compute $\alpha$ and the volume as a function of  $\overline{r}$ for different values of the contact angle. 

\subsection{Numerical results}
In Fig.~\ref{fig:alpha_against_r} we plot the direct dependence of the   folding angle with the droplet curvature,   $\overline{r}$, for three different values of the contact angle. When $\overline{r}$ tends to 0, $\alpha$ tends to ${\pi}/{2}$, and there is no folding. This is the large-stiffness limit. Inversely, when $\overline{r}$ tends to $+\infty$, $\alpha$ asymptotes to the value ${\pi}/{2}-\theta_E$. That value simply arises from the fact that the liquid-air interface is flat, and thus the sum of the folding angle and the contact angle needs to add up to $\pi/2$. Note that in this small-stiffness limit, the capillary pressure is zero (no curvature of the free surface) and the only force affecting the sheet is the surface tension acting locally on both contact lines.  As can be seen, both the hydrophilic and hydrophobic cases lead to folding of the elastic sheet, hydrophilic droplets leading to $\alpha > 0$ while the hydrophobic situation leads to  $\alpha < 0$.
Note that we have ignored here steric effects between both ends of the sheet (physical overlap), which would have to be addressed for all cases in which $\alpha<0$. 

\begin{figure}[t]
\includegraphics[width=.5\textwidth]{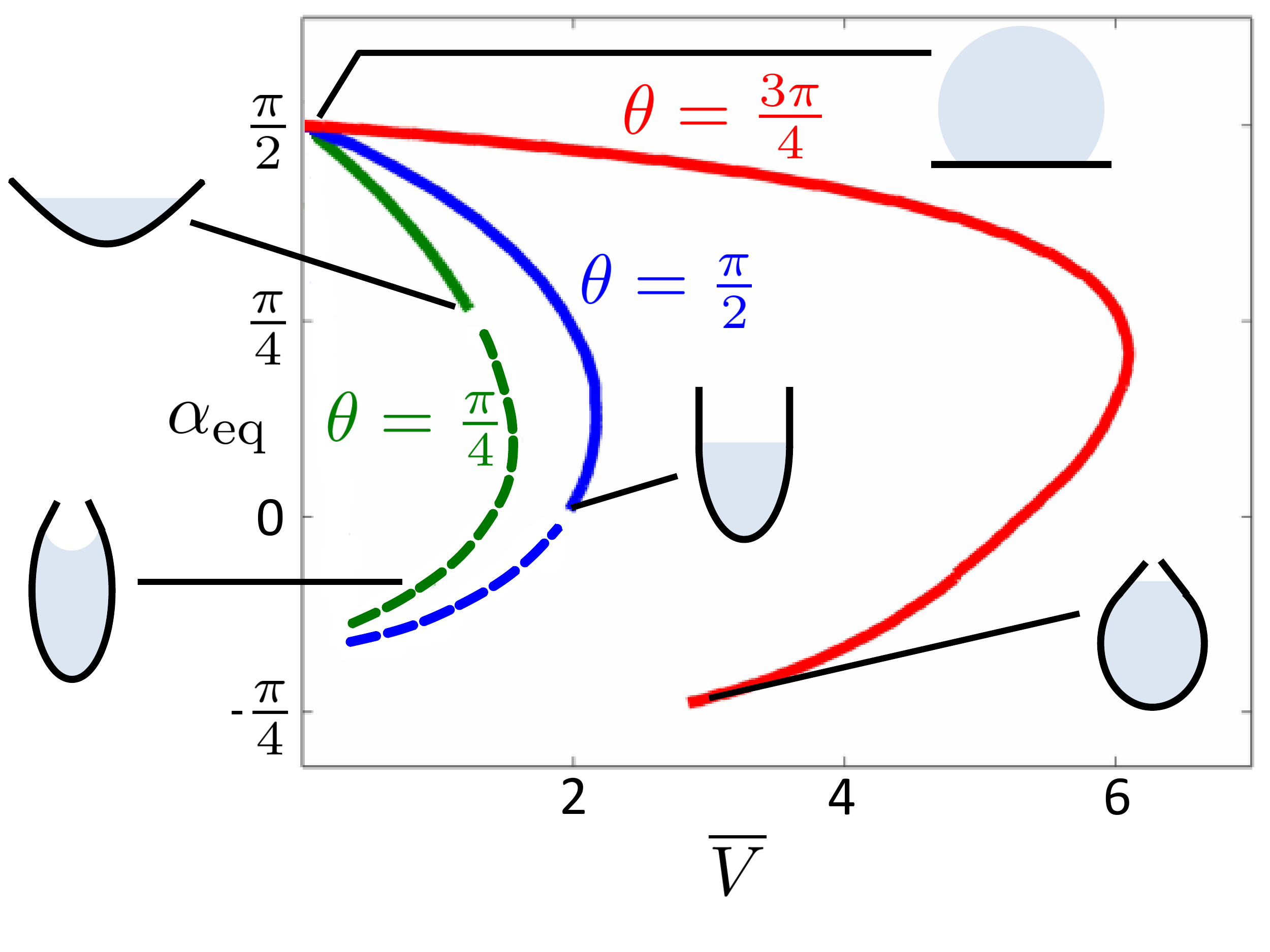}
\caption{Folding angle, $\alpha$, as a function of  the dimensionless droplet volume,  $\overline{V}=V{\gamma}/{B}$, for three values of the equilibrium contact angle ($\pi/4, \pi/2$, and $3\pi/4$). The dashed lines represent cases where $\overline{r}<0$.}
\label{fig:data_volume}
\end{figure}

After calculating the volume for each configuration, we can  plot the dependence of the folding angle on the dimensionless droplet volume,  $\overline{V}=V{\gamma}/{B}$. In order to compute the complete curve, we need to consider also negative values for $\overline{r}$ corresponding to the situation in which the droplet is  concave. The final results are shown in  Fig.~\ref{fig:data_volume} for three different values of the contact angle. Solid lines show cases with $r>0$ while dashed lines indicate $r<0$. For each contact angle, there exists a range of droplet volumes in which two solutions for $\alpha$ exist. In that case, the equilibrium  reached experimentally will depend on the initial droplet/sheet configuration. Given the typical  protocols discussed in the introduction, we expect that experiments will start in a completely unfolded geometry ($2\alpha=\pi$) and thus the equilibrium reached for a given volume in Fig.~\ref{fig:data_volume} will be the largest of the two values of $\alpha$.

\section{Conclusion}\label{discussion}

In this work, we have used theoretical calculations, scaling analysis, and numerical computations to address the role of geometry and wetting  on capillary  folding. Our study was motivated by a number of experiments  demonstrating the folding of two-dimensional templates into complex three-dimensional shapes, and the goal of the paper was to provide an overview of the various experimental control parameters. 

One of the  important messages of our paper is the demonstration that by simply varying the contact angle between the liquid-air interface and the solid surface, the complete range of folding angles can be obtained, from complete folding ($\alpha=0$) to a fully flat configuration ($\alpha=\pi/2$). The key formula,  derived in \S \ref{single_droplet} and seen repeatedly throughout the paper, relates the folding angle, $\alpha_\text{eq}$, to the equilibrium contact angle, $\theta_{E}$, as
\begin{equation}\label{main}
\alpha_\text{eq}=-\frac{\pi}{2} + \theta_{E}.
\end{equation}
This equation governs the equilibrium in the case of the infinite walls, without de-wetting, and a hydrophobic droplet. Even in geometrically complicated situations such as in the presence of non-straight walls or with contact angle hysteresis, Eq.~(\ref{main}) can still be used provided the parameters $\alpha$ and $\theta$ are defined correctly, as detailed above.

In \S \ref{2D}, the idealized two-dimensional geometry allowed us to  consider  different  parameters allowing to  control the folding angle.  The volume of the droplet is one of these tuning parameters, either  when the droplet is large enough that it reaches the edges of the walls (\S \ref{finite-size}), when the walls display nontrivial shapes (\S \ref{curved} and \ref{kinks}), or when we add another droplet on the other side of the ridge (\S \ref{2drop}). The ability for a droplet to dewet from the corner of the folded region or the presence of contact angle hysteresis  can have major consequences too (\S\ref{dewet}). If dewetting  theoretically leads to an undetermined minimum energy configuration, it appears  that contact angle hysteresis   would prevent  dewetting from occurring (\S \ref{hyst}).  These results, derived theoretically in two dimensions, appear to remain valid in three dimensions  (\S \ref{3D}). We finally considered and solved the problem of capillarity-driven folding of an elastic sheet in two dimensions (\S \ref{elastic}). Once again, the contact angle and the volume of the droplet were important control parameters, along with flexibility. 

One of the exciting avenues for extending this line of work would be to address the inverse folding problem. We now have a thorough physical picture of how geometrical and material parameters affect folding in two and three dimensions. Would it be then possible to design the two-dimensional pattern and control  parameters  able to yield a particular folded three-dimensional structure? Can  shapes of arbitrary complexity be obtained or are there intrinsic limits to the use of surface tension for small-scale manufacturing? We hope that our study will motivate further work along these directions.

\section*{Acknowledgments}

This work was funded in part by the NSF (grant CBET CBET-0746285 to Eric Lauga).

\bibliographystyle{unsrt}
\bibliography{folding_bib}
\end{document}